\documentclass[fleqn,usenatbib]{mnras}

\usepackage{newtxtext,newtxmath}

\usepackage[T1]{fontenc}
\usepackage{ae,aecompl}

\usepackage{graphicx}	% Including figure files
\usepackage{amsmath}	% Advanced maths commands
\usepackage{placeins}
\usepackage{multirow}
\usepackage{xcolor}

\newcommand{\orcid}[1]{\textsuperscript{\,\,\href{https://orcid.org/#1}{\includegraphics[scale=0.06]{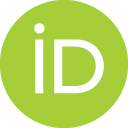}}}}

\newcommand{\HA}{H$\alpha$}
\newcommand{\NH}{[{N\sc{ii}}]/H$\alpha$}
\newcommand{\vsig}{v/$\sigma$}
\newcommand{\rh}{$R_{\rm h}$}
\newcommand{\med}[1]{#1}

\newcommand{\NII}{[N{\sc{ii}}]}

\newcommand{\zra}[2]{\mbox{$z$\,$\approx$\,#1\,--\,#2}}

\newcommand{\SFR}{M$_{\odot}\,$yr$^{-1}$}
\newcommand{\eagle}{{\sc{eagle}}}
\newcommand{\vsigma}{$v_{\rm rot,H\alpha}$/$\sigma_{\rm H\alpha}$}

\newcommand{\changed}[1]{{\color{black}#1}}
\title[Gas-Phase Metallicity of SFGs at $z$\,$\approx$\,0.6\,--\,1.8]{The Evolution of Gas-Phase Metallicity and  Resolved Abundances in Star-forming Galaxies at $z$\,$\approx$\,0.6\,--\,1.8}

\author[S. Gillman et al.]{S. Gillman\orcid{0000-0001-9885-4589},$^{\hyperlink{CEA}{1},\hyperlink{DAWN}{2},\hyperlink{DTU}{3}}$\thanks{E-mail: srigi@space.dtu.dk}
A. L.  Tiley\orcid{0000-0002-0617-9510},$^{\hyperlink{CEA}{1},\hyperlink{ICRAR}{4}}$
A. M. Swinbank\orcid{0000-0003-1192-5837},$^{\hyperlink{CEA}{1}}$
U. Dudzevi\v{c}i\={u}t\.{e}\orcid{0000-0003-4748-0681},$^{\hyperlink{CEA}{1}}$
\and
R. M. Sharples\orcid{0000-0003-3449-8583},$^{\hyperlink{CEA}{1},\hyperlink{CFAI}{5}}$
Ian Smail\orcid{0000-0003-3037-257X},$^{\hyperlink{CEA}{1}}$
C. M. Harrison\orcid{0000-0001-8618-4223},$^{\hyperlink{Newcastle}{6}}$
Andrew J. Bunker\orcid{0000-0002-8651-9879},$^{\hyperlink{Oxford}{7},\hyperlink{Japan}{8}}$
\and 
Martin Bureau\orcid{00000-0003-4980-1012},$^{\hyperlink{Oxford}{7}}$
M. Cirasuolo\orcid{0000-0003-1644-8881},$^{\hyperlink{ESO}{9}}$
Georgios E. Magdis\orcid{0000-0002-4872-2294},$^{\hyperlink{DAWN}{2},\hyperlink{DTU}{3},\hyperlink{Copenhagen}{10},\hyperlink{Greece}{11}}$
\and  Trevor Mendel\orcid{0000-0002-6327-9147}$^{\hyperlink{ANU}{12}}$
and John P. Stott\orcid{0000-0002-1679-9983}$^{\hyperlink{Lancaster}{13}}$
\\
$^{1}$\hypertarget{CEA}{Centre for Extragalactic Astronomy, Durham University, South Road, Durham, DH1 3LE UK}\\
$^{2}$\hypertarget{DAWN}{Cosmic Dawn Center (DAWN), Copenhagen, Denmark}\\
$^{3}$\hypertarget{DTU}{DTU-Space, Technical University of Denmark, Elektrovej 327, DK-2800 Kgs. Lyngby, Denmark}\\
$^{4}$\hypertarget{ICRAR}{International Centre for Radio Astronomy Research, University of Western Australia, 35 Stirling Highway, Crawley, WA 6009, Australia}\\
$^{5}$\hypertarget{CFAI}{Centre for Advanced Instrumentation, Durham University, South Road, Durham DH1 3LE UK}\\
$^{6}$\hypertarget{Newcastle}{School of Mathematics, Statistics and Physics, Newcastle University, Newcastle upon Tyne NE1 7RU, UK}\\
$^{7}$\hypertarget{Oxford}{Sub-department of Astrophysics, Department of Physics, University of Oxford, Denys Wilkinson Building, Keble Road, Oxford OX1 3RH, UK}\\
$^{8}$\hypertarget{Japan}{Kavli Institute for the Physics and Mathematics of the Universe (WPI), The University of Tokyo, Kashiwa, Chiba 277-8583, Japan}\\
$^{9}$\hypertarget{ESO}{European Southern Observatory, Karl-Schwarzschild-Str 2, D-86748 Garching b. M{\"u}nchen, Germany} \\
$^{10}$\hypertarget{Copenhagen}{University of Copenhagen, Lyngbyvej 2, DK-2100 Copenhagen \O, Denmark}\\
$^{11}$\hypertarget{Greece}{Institute for Astronomy, Astrophysics, Space Applications and Remote Sensing, National Observatory of Athens, GR-15236 Athens, Greece}\\
$^{12}$\hypertarget{ANU} {Research School of Astronomy and Astrophysics, Australian National University, Canberra, ACT 2611, Australia}\\
$^{13}$\hypertarget{Lancaster}{Department of Physics, Lancaster University, Bailrigg, Lancaster LA1 4YB, UK}\\
}

\date{Accepted 2020 October 28. Received 2020 October 19; in original form 2020 August 28}

\pubyear{2020}

\begin{document}
\label{firstpage}
\pagerange{\pageref{firstpage}--\hyperlink{lp}{20}}%\pageref{lastpage}}
\maketitle

\begin{abstract}
We present an analysis of the chemical abundance properties of $\approx$650 star-forming galaxies at
\zra{0.6}{1.8}. Using integral-field observations from the $K$\,-\,band Multi-Object Spectrograph (KMOS), we quantify the \NH{} emission-line ratio, a proxy for the gas-phase Oxygen abundance within the interstellar medium. We define the stellar mass\,--\,metallicity relation at \zra{0.6}{1.0} and \zra{1.2}{1.8} and analyse the correlation between the scatter in the relation and fundamental galaxy properties (e.g. \HA{} star-formation rate, \HA{} specific star-formation rate, rotation dominance, stellar continuum half-light radius and Hubble-type morphology). We find that for a given stellar mass, more highly star-forming, larger and irregular galaxies have lower gas-phase metallicities, which may be attributable to their lower surface mass densities and the higher gas fractions of irregular systems. We measure the radial dependence of gas-phase metallicity in the galaxies, establishing a median, beam smearing-corrected, metallicity gradient of
\med{$ \Delta Z / \Delta R $}=\,0.002\,$\pm$\,0.004 dex kpc$^{-1}$, indicating on average there is no significant dependence on radius. The metallicity gradient of a galaxy is independent of its rest-frame optical morphology, whilst correlating with its stellar mass and specific star-formation rate, in agreement with an inside-out model of galaxy evolution, as well as its rotation dominance. We quantify the  evolution of metallicity gradients, comparing the distribution of $\Delta Z / \Delta R$ in our sample with numerical simulations and observations at \zra{0}{3}. Galaxies in our sample exhibit flatter metallicity gradients than local star-forming galaxies, in agreement with numerical models in which stellar feedback plays a crucial role redistributing metals.
\end{abstract}

\begin{keywords}
galaxies: abundances -- galaxies: high-reshift -- galaxies: kinematics and dynamics
\end{keywords}

%%%%%%%%%%%%%%%%%%%%%%%%%%%%%%%%%%%%%%%%%%%%%%%%%%

%%%%%%%%%%%%%%%%% BODY OF PAPER %%%%%%%%%%%%%%%%%%

\section{Introduction}

The ejection of metals into the interstellar medium via winds from massive stars in the asymptotic giant branch phase or via supernovae acts to increase the chemical abundance in star-forming galaxies. This influx of metals is mediated by inflows, outflows and cold gas accretion \citep[e.g.][]{Koppen1999,Calura2006,Erb2008,Steidel2010,Dave2012,Lu2015,Sanders2015,Angle2017,Christensen2018,Sanders2018}. Empirically constraining the complex interplay between these secular processes and their imprint on the chemical abundance properties of galaxies is crucial to fully constrain the baryon cycle.

In the local Universe, the correlation between a galaxy's stellar mass ($M_{*}$) and its integrated gas-phase metallicity ($ Z$), the mass\,--\,metallicity relation, has been well studied \citep[e.g.][]{Lequeux1979,Tremonti2004}. Higher stellar mass  star-forming galaxies have been shown to exhibit higher gas-phase metallicities at $z$\,$\approx$\,0 with $Z\,\propto\,M_{*}^{2/5}$ up to $M_{*}$\,$\sim$\,10$^{10}$M$_{\odot}$, above which the relation saturates to a constant metallicity. This strong correlation between stellar mass and metallicity is believed to be a consequence of supernovae driven winds and outflows  which remove the metal-rich gas from the interstellar medium as well as the inflow of metal-poor inter-galactic medium gas \citep[e.g.][]{Tremonti2004,Tumlinson2011,Dayal2013}. In lower stellar mass galaxies, with shallower potential wells, a larger fraction of this material is removed from the galaxy resulting in an overall lower metallicity \citep[e.g.][]{Arimoto1987,Garnett2002,Brooks2007,Dayal2013,Chisholm2018}. Higher stellar mass galaxies are also believed to evolve more rapidly at higher redshifts and have therefore converted more of their pristine gas into stars and metals, resulting in higher gas-phase metallicities \citep[e.g.][]{Maiolino2008,Somerville2015,Sanders2018}.

The star-formation rate and gas fraction of a galaxy have also been linked to its offset from the mass\,--\,metallicity relation, with suggestions of the existence of a fundamental metallicity plane. In this plane, more highly star-forming  galaxies, at a given stellar mass, have lower gas-phase metallicities \citep[e.g.][]{Mannucci2010,Sanchez2017,Sanchez-Menguiano2019}. Surveys of galaxies in the local Universe, such as Calar Alto Legacy Integral Field Area Survey \citep[CALIFA;][]{Sanchez2012} and Mapping Nearby Galaxies at Apache Point Observatory \citep[MaNGA;][]{Bundy2015}, which observed thousands of galaxies out to $z$\,$\approx$\,0.03, have shown a correlation between galaxy star-formation rate and offset from the  mass\,--\,metallicity  relation \citep[e.g.][]{Sanchez2019,Cresci2019}.

Attempts have been made to define the mass\,--\,metallicity  relation (and plane) in the distant Universe. The ratio of strong optical nebular metal emissions lines to the Balmer series  (e.g. \NH{}) is commonly used to infer the gas-phase metallicity of galaxies at high redshift due to their spectral proximity, making them insensitive to dust and observable from ground-based facilities \citep[e.g.][]{Yabe2015,Wuyts2016,Schreiber2018,Curti2019}. 

The ratio is often expressed as an Oxygen abundance relative to Hydrogen, as Oxygen is generally the most abundant heavy element by mass and therefore provides a proxy for the metallicity of the galaxy \citep{Pettini2004}. Using these strong-line calibrations of optical emission lines, many recent studies have shown that the mass\,--\,metallicity  relation evolves above $z$\,=\,1, with higher-redshift galaxies having lower metallicities at a given stellar mass \citep[e.g.][]{Erb2006,Maiolino2008,Zahid2011,Stott2013b,Yabe2015,Wuyts2016,Sanders2018}.

Despite evolution in the normalisation of the mass\,--\,metallicity  relation, it has been suggested that the fundamental plane of stellar mass, gas-phase metallicity and star-formation rate does not evolve with cosmic time. For example \citet{Mannucci2011} proposed that the evolution in the mass\,--\,metallicity  relation comes from sampling different regions of the non-evolving fundamental plane, given the higher average star formation at earlier cosmic times \citep{Cresci2019}. To confirm this result, metallicity measurements of large samples of high-redshift galaxies are required, which has recently been made possible with the advent of high-redshift multi-object spectroscopy. 

Using the MOSFIRE Deep Evolution Field (MOSDEF) Survey, \citet{Sanders2015,Sanders2018,Sanders2020} demonstrated the presence of the \changed{relation between stellar mass, metallicity and star\,--\,formation rate} in 300 galaxies at $z$\,$\approx$\,2.3 and 150 galaxies at $z$\,$\approx$\,3.3. Star-forming galaxies in the MOSDEF survey exhibit metallicities within 0.04\,dex of local galaxies at fixed stellar mass and star-formation rate. \citet{Sanders2020} concludes that there is no evidence that the fundamental plane of stellar mass, gas-phase metallicity and star-formation evolves out to $z$\,$\approx$\,3.3, with the lower metallicity at earlier cosmic times, for a given stellar mass, being driven by higher gas fractions and higher metal removal efficiency.

As well as the galaxy integrated metallicity, the distribution of metals within a galaxy provides insights into the influence of  star formation, gas accretion, mergers and feedback, that all play a key role in defining the evolution of galaxies. In the local Universe most isolated galaxies exhibit negative log-linear abundance gradients (i.e metallicity decreasing with radius). Higher gas-phase metallicities are observed in the central regions of the galaxies  where star formation is most prevalent and supernovae enrich the surrounding interstellar medium \citep[e.g.][]{Sanchez2014,Kaplan2016,Poet2018}. The observed negative gradients are predicted from inside-out theories for the growth of galaxy discs \citep[e.g.][]{Boisser1999} and are well modelled in hydrodynamical simulations such as Evolution and Assembly of GaLaxies and their Environments (\eagle{}) \citep[e.g.][]{DeRossi2017,Collacchioni2020}.

\begin{figure*}
    \centering
    \includegraphics[width=\linewidth,trim={0cm 2.5cm 0cm 0cm}]{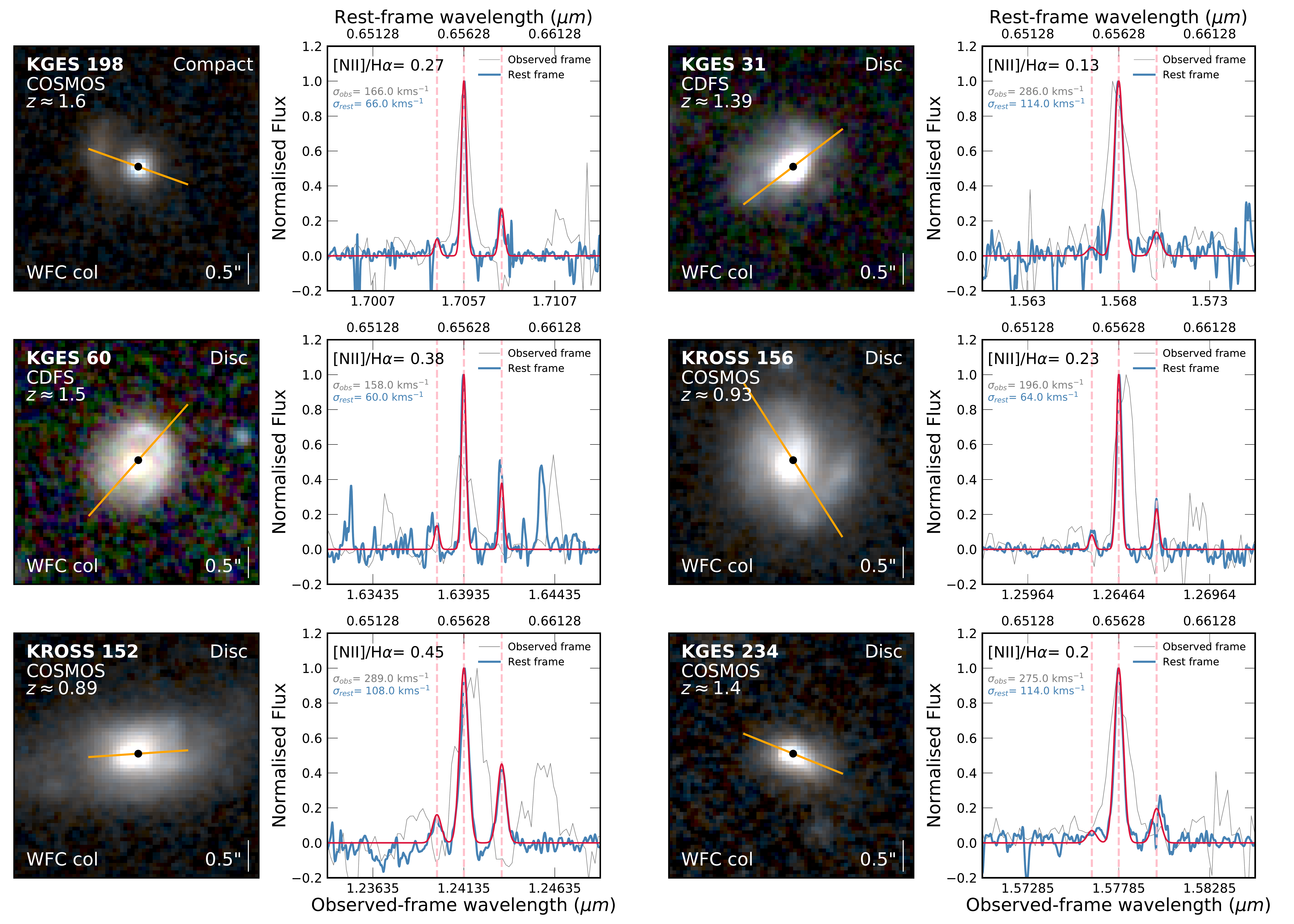}
    \caption{Example $HST$ images and integrated spectra of galaxies in our sample. For each galaxy we show a WFC3 three-colour image composed of F105W, F125W and F160W images. The semi-major axis of the galaxy (orange line) and stellar continuum centre (black filled circle) are indicated, as well as the morphological class if available from the \citet{Huertas-Company2015} classification. The galaxy integrated spectra from the observed KMOS data cube (grey) and de-redshifted rest-frame cube with bulk motions removed (blue) are also shown. The \HA{} and \NII{} emission lines are indicated (vertical pink dashed lines) and we overlay the spectral fit to the de-redshifted and corrected emission lines (crimson line). The \NH{} ratio of each galaxy, \changed{as well as the velocity dispersion of the spectra are} indicated in the top left. The examples show a range of \NH{} ratio from galaxies with varying morphologies and redshifts.}
    \label{KGES_metal_fig:zp_examples}
\end{figure*}

At high redshift the picture is much less clear, with various observational studies reporting a range of metallicity gradients in isolated star-forming galaxies \citep[e.g.][]{Cresci2010,Jones2010,Swinbank2012a,Jones2013,Stott2014,Leethochawalit2016,Wuyts2016,Molina2017, Wang2017,Curti2019} as well as in simulations \citep[e.g][]{Pilkington2012,Angle2014,Finlator2017,Sillero2017,Collacchioni2020,Hemler2020}.
In this paper, we exploit a large sample of high-redshift star-forming galaxies with spatially-resolved metallicity profiles. Utilizing the strong-line calibration of the \NH{} ratio, we present an analysis of the gas-phase metallicity properties of $\approx$\,650 star-forming galaxies in the redshift range \zra{0.6}{1.8}. We explore the correlations between galaxy morphology, dynamics and position on the mass\,--\,metallicity relation, as well as the metallicity profile of the galaxies.

This paper is organized as follows. In Section \ref{KGES_metal_sec:sample} we describe the observations and sample selection. We define the procedure used to correct for the galaxies velocity fields and de-redshift the integral-field data, from which we extract the gas-phase metallicities. In Section \ref{KGES_metal_sec:MZR} we present the mass\,--\,metallicity  relation and fundamental metallicity plane of the sample. We analyse the radial dependence of metallicity and its correlation with galaxy dynamics, morphology and redshift in Section \ref{KGES_metal_Sec:zgrad}, before presenting our conclusions in Section \ref{KGES_metal_Sec:Conc}.

A nine-year {\it{Wilkinson Microwave Anisotropy Probe}} \citep{Hinshaw2013} cosmology is used throughout this work with $\Omega_{\Lambda}$\,=\,0.721, $\Omega_{ m}$\,=\,0.279 and $H_{\rm 0}$\,=\,70\,km\,s$^{-1}$ Mpc$^{-1}$.
In this cosmology an angular resolution of 0.70 arcsecond (the median full width half maximum (FWHM) of the seeing in our data) corresponds to a physical scale of 5.5\,kpc at a redshift $z$\,=\,0.91 (the median redshift of our data). All quoted magnitudes are in the AB system and stellar masses are calculated assuming a Chabrier initial mass function (IMF) \citep{Chabrier2003}.

\begin{figure*}
    \centering
    \includegraphics[width=\linewidth,trim={0cm 1cm 0cm 0cm}]{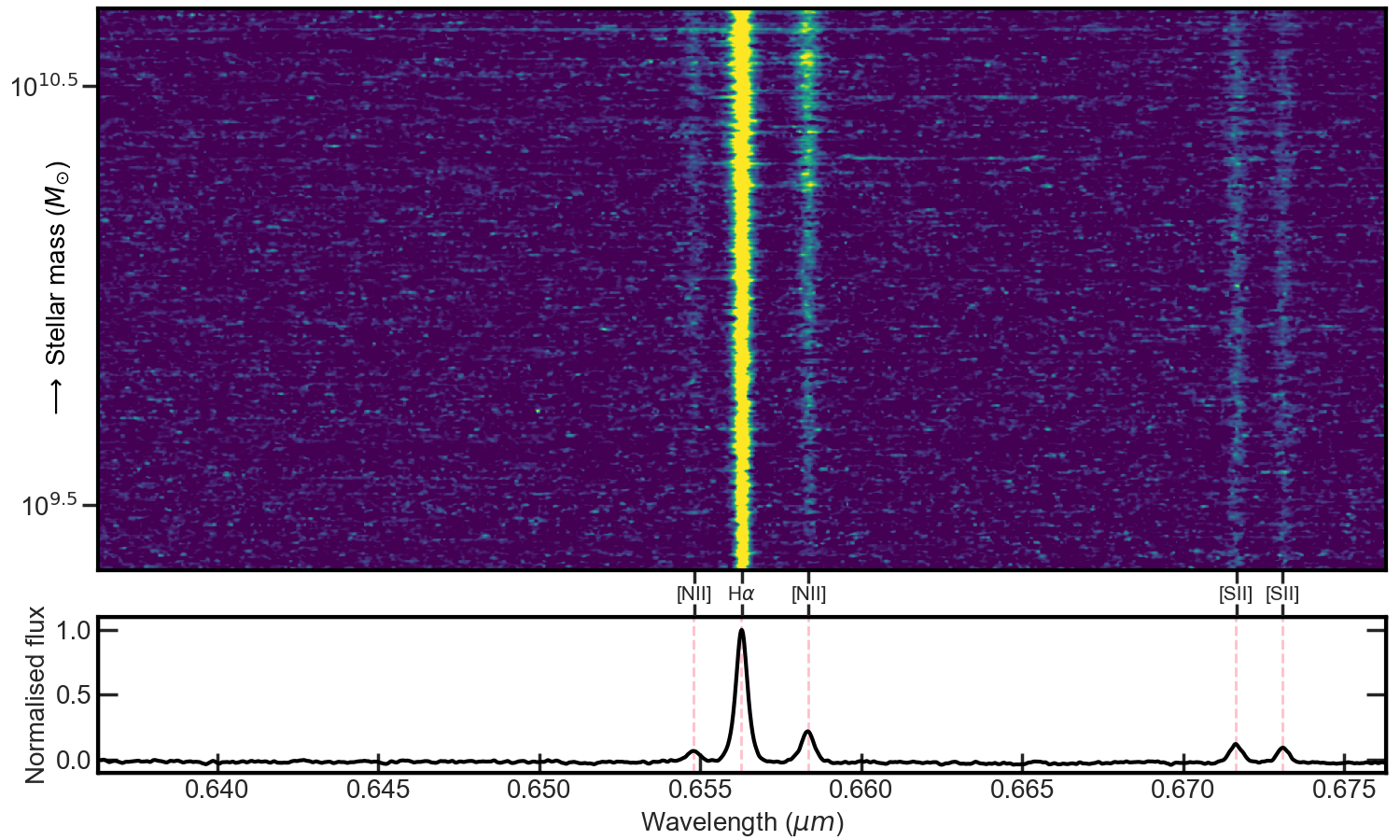}
    \caption{\textbf{Top}: Spectra of all 644 spatially-resolved KMOS galaxies, with the de-redshifting and velocity correction procedure applied, ranked by stellar mass. The spectra of the higher stellar mass galaxies reveal the weaker [{N\sc{ii}}] forbidden emission lines whilst the stronger recombination emission line (H$\alpha$) and [S{\sc{ii}}] forbidden emission line are present in all galaxies. This reflects the increase of the gas-phase metallicity with stellar mass seen in the mass\,--\,metallicity relation. \textbf{Bottom}: Normalised stacked spectrum over all 644 KMOS observations, with both forbidden and recombination emission lines indicated (vertical pink dashed lines).}
    \label{KGES_metal_fig:KMOS_spectra}
\end{figure*}

\section{Sample Selection and Analysis}\label{KGES_metal_sec:sample}

To provide statistically meaningful conclusions about the metallicities of galaxies in the distant Universe, we utilize 644 star-forming galaxies drawn from two large $K$-band multi-object spectrograph (KMOS; \citealt{Sharples2004,Sharples2013}) programmes at \zra{0.6}{1.8}. The galaxies in our sample are drawn from the KMOS Redshift One Spectroscopic Survey (KROSS; \citealt{Stott2016,Harrison2017}) at \zra{0.6}{1.0} (448 galaxies) and the KMOS Galaxy Evolution Survey (KGES; \citealt{Gillman2019b}, Tiley et al. in prep.) at \zra{1.2}{1.8} (196 galaxies).

All the galaxies in the sample were observed using KMOS, a multi-object spectrograph mounted on the Nasmyth focus of the 8-m class UT1 telescope at the VLT, Chile. It has 24 individual integral-field units that patrol a 7.2 arcminute diameter field, each with a 2.8\,$\times$\,2.8 arcsecond$^{2}$ field of view and 0.2\,$\times$\,0.2 arcsecond$^{2}$ spaxels. KMOS utilizes image slices to produce data cubes with wavelength coverages of 0.8\,--\,1.0, 1.0\,--\,1.4, 1.4\,--\,1.9, 1.9\,--\,2.5 or 1.5\,--\,2.5\,$\mu$m in the $IZ$, $YJ$, $H$, $K$ or $HK$ band respectively. In the following sections we provide an overview of the KROSS and KGES surveys.

\subsection{KROSS}
KROSS is a KMOS survey of 795 typical star-forming galaxies in the redshift range \zra{0.6}{1.0} selected from the Extended $Chandra$ Deep Field South (E-CDFS; \citealt{Giacconi2001}), Cosmological Evolution Survey (COSMOS; \citealt{Scoville2007}), UKIRT Infrared Deep Sky Survey (UKIDSS) Ultra-Deep Survey (UDS; \citealt{Lawrence2007}) and the SA22 \citep{Lilly1991,Steidal1998} extragalactic fields. The majority of the sample galaxies are selected using archival spectroscopic redshifts, 25 per cent being H$\alpha$ narrow-band emitters at $z$\,=\,0.84 from the High Redshift Emission Line Survey (HiZELS) and  Canada-France HiZELS (CF-HIZELS) surveys \citep{Sobral2013,Sobral2015}. A full description of the KROSS survey and derived galaxy properties is given in \citet{Stott2016} and \cite{Harrison2017}.

The KROSS targets were predominantly selected to have bright H$\alpha$ emission in the $J$-band, avoiding OH skylines, and a total apparent magnitude $K$\,$<$\,22.5, with a `blue' colour $r$\,--\,$z$\,$<$\,1.5. Of the 795 galaxies targeted, 586 were detected in H$\alpha$ emission. Removing galaxies with \NH{}\,$>$\,0.8 and/or a $\gtrsim$\,1000 \,km\,s$^{-1}$ broad-line component in the \HA{} emission-line profiles, indicating the presence of an active galactic nucleus (AGN), leaves 448 galaxies, for which we can measure the gas-phase metallicities. These 448 galaxies of the KROSS sample have a median stellar mass of \med{$\log(M_*[M_{\odot}])$}\,=\,10.0 with a 16\,--\,84th percentile range of $\log(M_*[M_{\odot}])$\,=\,9.6\,--\,10.4 as derived using the {\sc{hyperz}} \citep{HYPERZ2000} spectral energy distribution (SED) fitting code. The dust-corrected H$\alpha$ star-formation rates of the sample are derived following \citet{Kennicutt1998}, \changed{corrected to a Chabrier initial mass function and assuming a median \HA{} extinction, calculated from SED fitting, of $A_{\rm H\alpha}$\,=\,1.73. The \zra{0.6}{1.0} sample has a median value, and bootstrap uncertainty, of} \med{SFR}\,=\,7\,$\pm$\,1\,\SFR{} with a 16\,--\,84th percentile range of SFR\,=\,4\,--\,15\,\SFR{}(see \citealt{Stott2016} and \citealt{Harrison2017} for details).

\subsection{KGES}
KGES is a KMOS survey that targets 288 star-forming galaxies at \zra{1.2}{1.8} that preferentially lie within some of the Cosmic Assembly Near-infrared Deep Extragalactic Legacy Survey \citep[CANDELS;][]{Grogin2011} \textit{Hubble Space Telescope} ($HST$) fields. 
The programme probes ongoing star formation in the galaxies using the H$\alpha$ and [{N\sc{ii}}] emission lines. Galaxies with a $K$-band magnitude of $K$\,$<$\,22.5 are selected from the UDS, COSMOS and E-CDFS extragalactic fields. No prior morphological selection was made, with the remaining KMOS integral field unit (IFU) arms filled with fainter galaxies. Of the 288 galaxies targeted, 243 were detected in H$\alpha$ emission and 235 have spatially-resolved H$\alpha$ emission. 

We remove potential AGN from the sample using the galaxies' infrared colours, following the \citet{Donley2012} and \citet{Stern2012} colour selection, and exclude galaxies with X-ray counterparts. We also remove galaxies with \NH>\,0.8 (see Tiley et al. in prep. for details) and galaxies with no \NII{} detection. This leaves 196 galaxies in which we can measure the integrated \NH{} ratio. These 196 galaxies of the KGES sample have a median stellar mass of \med{$\log(M_*[M_{\odot}])$}\,=\,10.0 with a 16--84th percentile range of $\log(M_*[M_{\odot}])$\,=\,9.6\,--\,10.6 as derived using the {\sc{magphys}} SED fitting code \citep{Gillman2019b}. We note that the {\sc{magphys}} SED fitting code is different to the {\sc{hyperz}} code used for the KROSS galaxies, which can lead to up a 0.3\,dex systematic discrepancy in stellar mass estimates \citep[e.g.][]{Mobasher2015}. The dust-corrected H$\alpha$ star-formation rate \changed{are calculated following the methods of \citep{Wuyts2013} assuming a \citet{Calzetti1994} extinction law (see Tiley et al. in prep.). The median dust-corrected H$\alpha$ star-formation rate, and bootstrap uncertainty,} of the sample is \med{SFR}\,=\,20\,$\pm$\,1\,\SFR with a 16\,--\,84th percentile range of SFR\,=\,8\,--\,45\,\SFR. 

We combine both the KROSS and KGES samples to create a sample of 644 star-forming galaxies at \zra{0.6}{1.8} for which we can measure the gas-phase metallicities as traced by the \NH{} ratios. In Figure \ref{KGES_metal_fig:zp_examples} we show examples of the galaxies in our sample, highlighting the range of rest-frame optical morphologies and \NH{} ratios.

\subsection{De-redshifting and correcting for velocity gradient}

The integrated spectrum of a galaxy contains information about its intrinsic emission-line properties. Any emission or absorption feature in the spectrum is broadened by motions along the line-of-sight. However, with integral-field data it is possible to model the two-dimensional velocity field and correct for this dynamical broadening, as we describe in this section. 

For each spectrum within the datacube of a galaxy, we re-normalise the spectrum using the spectroscopic redshift of the galaxy such that the H$\alpha$ emission line of the continuum centre, determined from broadband imaging, is centred at the rest-frame wavelength of \HA{}, $\lambda$6563\,\AA. Using the H$\alpha$ velocity map, as derived in \citet{Harrison2017} and Tiley et al. (in prep.), we then shift each spaxel's spectrum  to correct for the mean velocity at that spaxel, estimated from the velocity field of the galaxy. We note that the velocity field is derived using an adaptive binning technique on the datacube with a spaxel \HA{} signal-to-noise (S/N) threshold of S/N$\geq$5 (see \citealt{Harrison2017}, Tiley et al. in prep. for details). This velocity field thus defines the aperture in which we later calculate the de-redshifted integrated spectrum of the galaxy. Any spaxels outside of the velocity map (i.e. S/N$<$5) are excluded as we do not have a velocity correction for these spaxels.

To confirm the de-redshifting process has removed the broadening of the emission lines due to the spaxel to spaxel variation in the \HA{} velocity in each galaxy, we measure the velocity dispersion of each galaxy's integrated spectrum before and after de-redshifting. We note we do not account for the line broadening due to the velocity gradient within each individual spaxel. The observed H$\alpha$ emission lines of the \zra{0.6}{1.0} galaxies have a median velocity dispersion of $\sigma_{\rm gal}$\,=\,162\,$\pm$\,4\,km\,s$^{-1}$ with a 16th\,--\,84th percentile range of $\sigma_{\rm gal}$\,=\,106\,--\,258\,km\,s$^{-1}$. At \zra{1.2}{1.8} the galaxies have a median velocity dispersion of $\sigma_{\rm gal}$\,=\,246\,$\pm$\,9\,km\,s$^{-1}$ with a 16th\,--\,84th percentile range of $\sigma_{\rm gal}$\,=\,165\,--\,381\,km\,s$^{-1}$. After de-redshifting, the median velocity dispersion of the \zra{0.6}{1.0} sample is $\sigma_{\rm gal}$\,=\,82\,$\pm$\,2\,km\,s$^{-1}$ with a 16th\,--\,84th percentile range of $\sigma_{\rm gal}$\,=\,61\,--\,127\,km\,s$^{-1}$ whilst the \zra{1.2}{1.8} galaxies have a median of $\sigma_{\rm gal}$\,=\,85\,$\pm$\,2\,km\,s$^{-1}$ with a 16th\,--\,84th percentile range of $\sigma_{\rm gal}$\,=\,66\,--\,114\,km\,s$^{-1}$. This demonstrates the effect of removing bulk motions within the galaxy (see Figure \ref{KGES_metal_fig:zp_examples}). 

We apply this de-redshifting procedure to all 644 KMOS data cubes in our sample. To measure the \NH{} ratio in each galaxy, we first generate an integrated spectrum by spatially collapsing the de-redshifted data cubes. In Figure \ref{KGES_metal_fig:KMOS_spectra} we show the integrated spectra of all galaxies ranked by stellar mass. The H$\alpha$ emission line and [S{\sc{ii}}] forbidden lines are prominent in all galaxies, whilst the [{N\sc{ii}}] forbidden transition lines  at $\lambda$6548\AA{} and $\lambda$6583\AA{} are more prominent in higher stellar mass galaxies. This reflects the increase of the gas-phase metallicity with stellar mass seen in the mass\,--\,metallicity relation \citep[e.g.][]{Tremonti2004,Mannucci2010}.

\section{Mass\,--\,Metallicity Relation}\label{KGES_metal_sec:MZR}

To measure the gas-phase metallicity in our sample, we utilize the \NH{} emission line ratio both as an integrated quantity and as a function of radius.

We model the H$\alpha$ and [{N\sc{ii}}] emission lines using three Gaussian profiles with fixed wavelength offsets and coupled FWHMs, following the same procedure as the emission-line fitting in \citet{Gillman2019b}. The fitting procedure uses a five parameter model with continuum, H$\alpha$ peak intensity, line center, line width, and \NH{} ratio. We apply this fitting procedure to all observed galaxies with an H$\alpha$ integrated line flux S/N\,$>$\,5. We derive a median \NH{} ratio for the \zra{0.6}{1.0} galaxies of \med{\NH}\,=\,0.25\,$\pm$\,0.01 with a 16th\,--\,84th percentile range of \NH\,=\,0.10\,--\,0.42. Whilst the \zra{1.2}{1.8} galaxies have a median value of \med{\NH}\,=\,0.19\,$\pm$\,0.02 with a 16th\,--\,84th percentile range of \NH\,=\,0.10\,--\,0.37. In the local Universe a galaxy's gas-phase metallicity correlates strongly with other fundamental properties (e.g. stellar mass, star-formation rate, morphology) \citep[e.g.][]{Tremonti2004,Ellison2008,Sanchez-Menguiano2019}. In the next section, we analyse these correlations in our high-redshift sample of 644 star--forming galaxies.

\begin{figure*}
\centering
\begin{minipage}{.5\textwidth}
%   \centering
    \includegraphics[width=\linewidth,trim={0cm 1.5cm 0cm 0cm}]{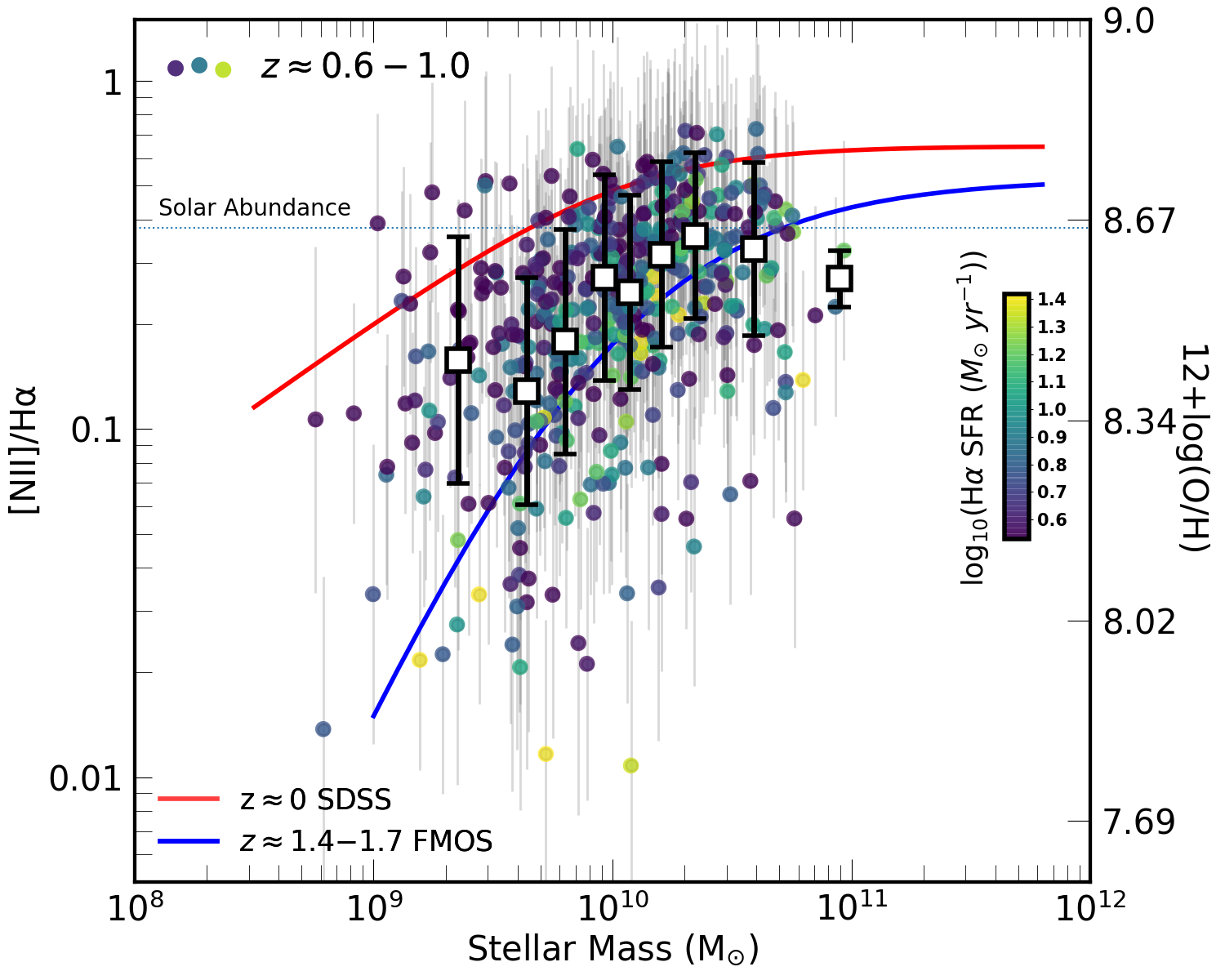}
\end{minipage}%
\begin{minipage}{.5\textwidth}
  \centering
 \includegraphics[width=\linewidth,trim={0cm 1.5cm 0cm 0cm}]{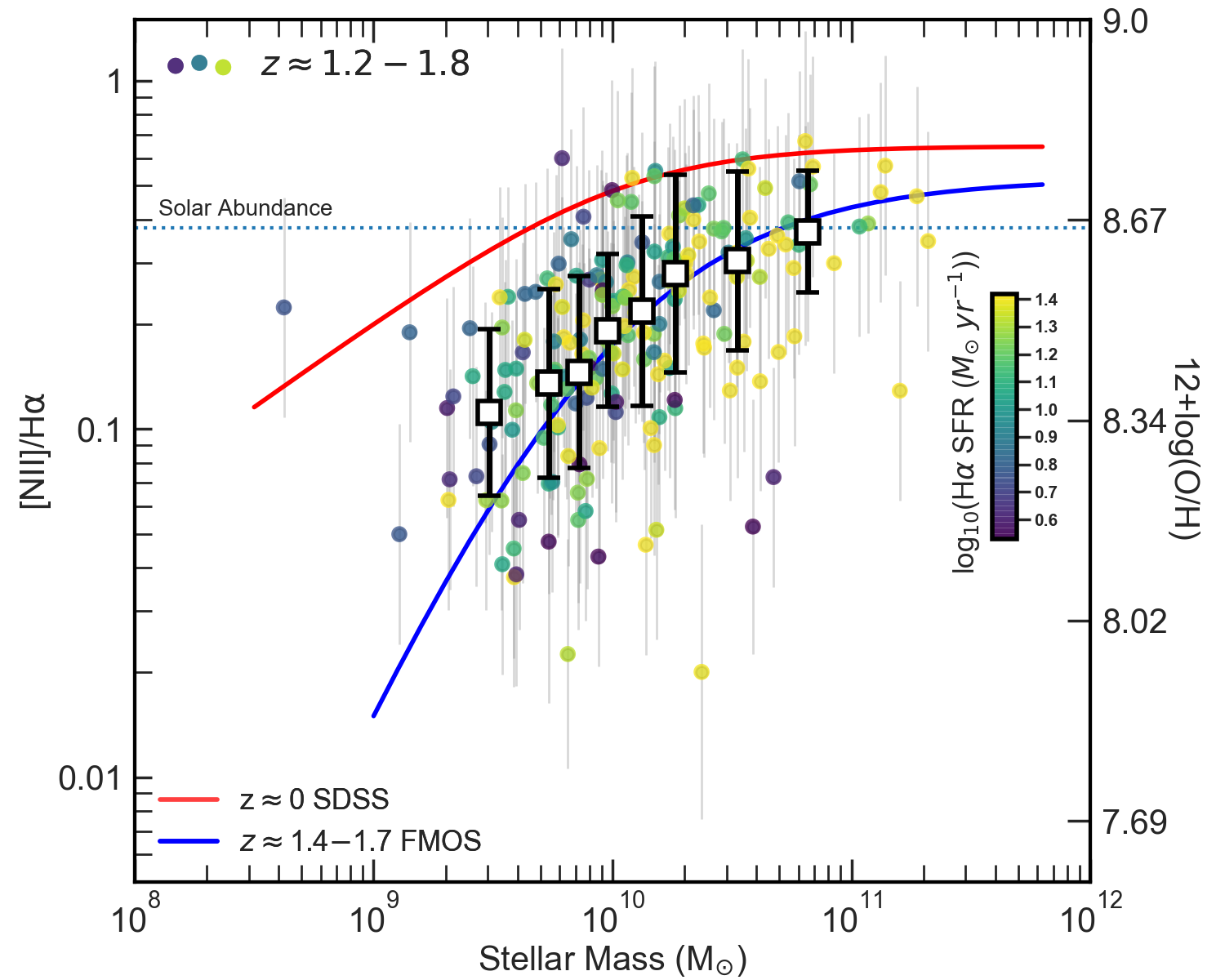}
\end{minipage}
\caption[]{Galaxy integrated gas-phase metallicity, measured from the integrated \NH{} ratio, as a function of stellar mass, coloured by the galaxy dust-corrected \HA{} star-formation rate, for both \zra{0.6}{1.0} galaxies (\textbf{left}) and \zra{1.2}{1.8} galaxies (\textbf{right}). The red line represents the mass\,--\,metallicity relation at $z$\,$\approx$\,0 derived from Sloan Digital Sky Survey (SDSS) \citep{Curti2020} whilst the blue line indicates the mass\,--\,metallicity relation at \zra{1.4}{1.70} from the Fiber-Multi Object Spectrograph (FMOS) survey \citep{Zahid2014}.
The horizontal dashed line indicates solar metallicity (12+log(O/H)\,=\,8.66; \citealt{Asplund2005}). The metallicity is converted from a \NH{} ratio to an Oxygen abundance following the linear calibration of \citet{Pettini2004}. The black squares (and error bars) indicate a running median and 1-$\sigma$ scatter\changed{, an interpolation of which is used to derive the offset from the median mass\,--\,metallicity relation ($\Delta Z$)}. In both redshift slices, higher stellar mass galaxies have higher metallicities, with an indication that more highly star-forming galaxies have lower gas-phase metallicities at a given stellar mass. The lower redshift \zra{0.6}{1.0} galaxies generally have higher metallicities than \zra{1.2}{1.8} galaxies, in agreement with the evolution between the \zra{1.4}{2.7} and $z$\,$\approx$\,0 observations. Note the limit \NH{}$<$0.8 at \zra{0.6}{1.0} and \zra{1.2}{1.8} to remove AGN, which are predominantly higher stellar mass galaxies.}
\label{KGES_metal_fig:NHratio_mstar}
\end{figure*}

\subsection{The mass\,--\,metallicity relation} \label{KGES_metal_subsec:mzr}

To quantify the processes that drive the baryon cycle within galaxies and ultimately drive their evolution, we analyse the connection between the metal content of the interstellar medium (e.g. gas-phase metallicity) and other fundamental galaxy properties such as stellar mass and star-formation rate.

To analyse the mass\,--\,metallicity relation at high redshift, we show the galaxy integrated \NH{} ratio as a function of stellar mass in Figure \ref{KGES_metal_fig:NHratio_mstar} for both the \zra{0.6}{1.0} and \zra{1.2}{1.8} samples. We convert the \NH{} ratio to an Oxygen abundance using the linear conversion from \citet{Pettini2004}, of the form, 
\begin{equation}
\rm     12 + \log (O/H) \,=\, 8.90 + 0.57\log([N{\textsc{ii}}] / H\alpha),
\end{equation}
which has a systematic uncertainty of 0.18\,dex. The median gas-phase metallicity of galaxies at \zra{0.6}{1.0} is  12+log(O/H)\,=\,8.56\,$\pm$\,0.01 with a 16\,--\,84th percentile range of 12+log(O/H)\,=\,8.34\,--\,8.69. For the \zra{1.2}{1.8} sub-sample the median metallicity is 12+log(O/H)\,=\,8.49\,$\pm$\,0.02 with a 16\,--\,84th percentile range of 12+log(O/H)\,=\,8.32\,--\,8.65. 

Higher stellar mass galaxies in our sample generally have higher metallicities, with a weaker dependence on stellar mass at the highest masses. This is in agreement with other high-redshift studies of galaxies \citep[e.g.][]{Stott2013b,Yabe2015,Wuyts2016,Schreiber2018,Sanders2018,Curti2020}, as well as inside-out galaxy evolution models and hydrodynamical simulations that predict feedback-driven winds and outflows which remove metal-rich material from lower-mass galaxies more easily than higher-mass systems, due to the their shallower potential wells \citep[e.g.][]{DeRossi2017,Angle2017,Chisholm2018,Gao2018}. 

In Figure \ref{KGES_metal_fig:NHratio_mstar} we overlay on our data trends from observations of low-redshift galaxies ($z$\,$\approx$\,0) from \citet{Curti2020} as well as high-redshift (\zra{1.4}{2.7}) galaxies from the Fiber-Multi Object  Spectrograph (FMOS) survey \citep{Zahid2014}. We note due to the different SED fitting codes used in the comparison samples there can be systematic offsets in stellar mass estimates (see \citealt{Mobasher2015}). On average the \zra{0.6}{1.0} sample is offset by 0.15\,$\pm$\,0.05\,dex to lower metallicities compared to the  $z \approx$\,0 star-forming galaxies from \citet{Curti2020}, whilst the \zra{1.2}{1.8} sample is offset by 0.22\,$\pm$\,0.05\,dex. Both samples have comparable metallicities to the \zra{1.4}{2.7} galaxies analysed by \citet{Zahid2014}. We note if we include galaxies with \NH{}>=0.8 the median trend at \zra{0.6}{1.0} increases on average by 0.03\,dex whilst the \zra{1.2}{1.8} median track is unaffected.

\begin{figure*}
    \centering
    \includegraphics[width=0.9\linewidth,trim={0cm 1cm 0cm 0cm}]{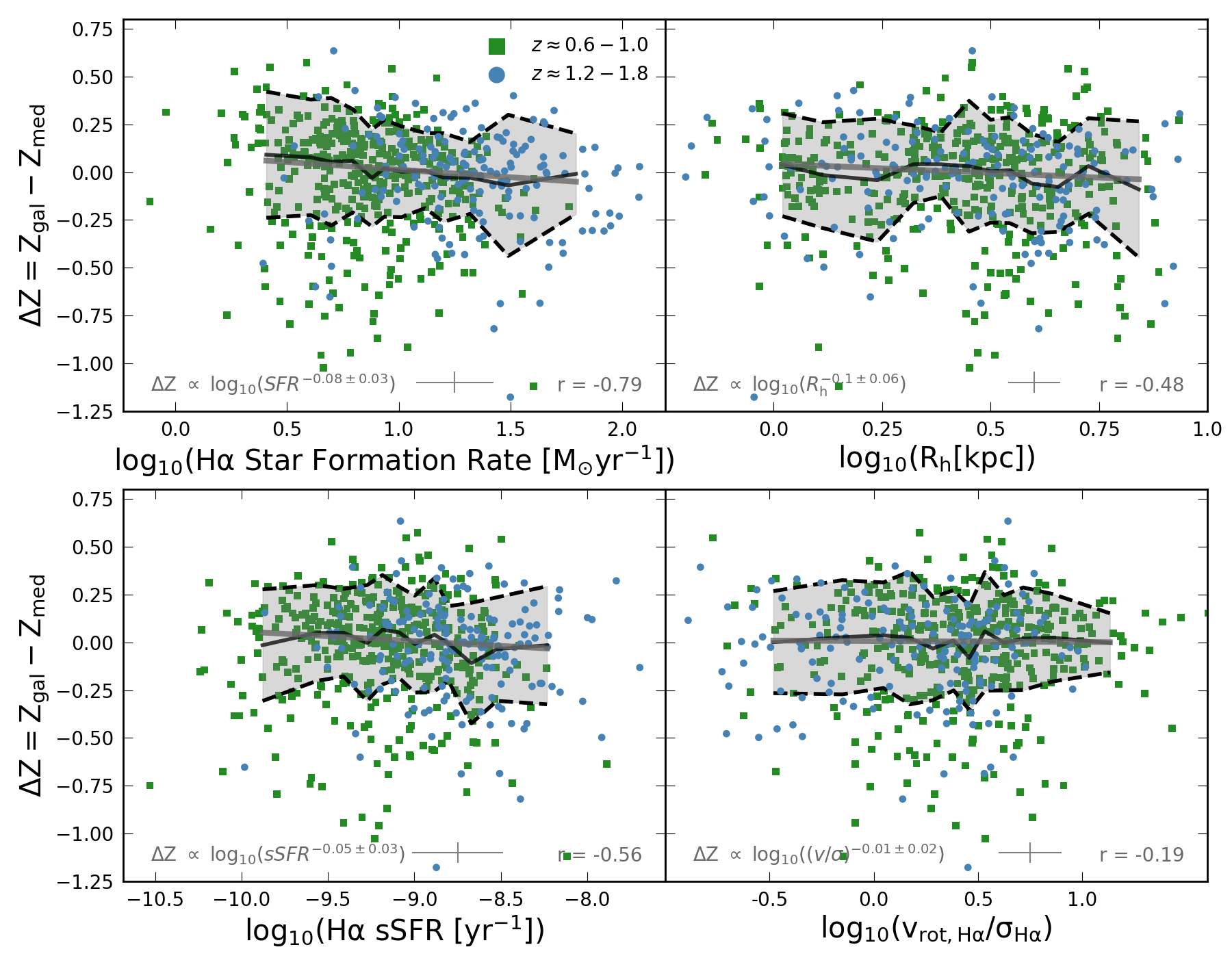}
    \caption[]{Offset from the median mass\,--\,metallicity relation ($\Delta Z$) as a function of H$\alpha$ star-formation rate (top-left panel), stellar continuum half-light radii (\rh{}) (top-right panel), H$\alpha$ specific star-formation rate (bottom-left panel) and \vsigma{} (top-right panel) for galaxies at \zra{0.6}{1.0}  and \zra{1.2}{1.8}.  \changed{The median mass\,--\,metallicity relation is derived by interpolating the running median shown in Figure \ref{KGES_metal_sec:MZR}}. The grey shaded region and black dashed lines indicate the 1\,--\,$\sigma$ scatter region on the running median (black solid line) and  in each panel. The grey line indicates a parametric fit to the running median, with parameters reported in the lower left corner of each panel. We also report the Spearman rank coefficient ($r$) of each correlation in the bottom-right corner of each panel. The median uncertainty is shown in the lower part of each panel.
    We identify a strong negative correlation between a galaxy's offset from the median mass\,--\,metallicity relation and its H$\alpha$ star-formation rate ($r$\,=\,--\,0.79) and a moderate correlation with the galaxy's specific star formation ($r$\,=\,--\,0.56). We find a moderate correlation between the offset from the median mass\,--\,metallicity relation and the stellar continuum half-light radius ($r$\,=\,--\,0.48) whilst the ratio of rotational velocity and velocity dispersion shows no correlation with a galaxies position in the mass\,--\,metallicity plane. This indicates that more highly star-forming, larger galaxies have a lower metallicity at a fixed stellar mass irrespective of their rotation dominance.}
    \label{KGES_metal_fig:dz_corr}
\end{figure*}

The gas-phase metallicities of our sample galaxies agrees with the evolution of the mass\,--\,metallicity relation identified by other surveys of star-forming galaxies \citep[e.g.][]{Huang2019,Sanders2020}. To determine whether the fundamental plane of stellar mass, gas-phase metallicity and star-formation evolves with redshift, we compare the median track of the \zra{0.6}{1.0} and \zra{1.2}{1.8} samples to that predicted by the  \citet{Curti2020} parameterisation of the plane.  The plane is parameterised as,
\begin{equation}\label{eq:FMR}
Z(M,SFR)\,=\,Z_{0}-(\gamma/\beta)\log(1+(M_*/M_0(SFR))^{-\beta})    
\end{equation}
where $\log(M_{0}(SFR))$\,=\,$m_0$+$m_1\log$(SFR). \citet{Curti2020} established the best-fitting parameters are $Z_0$\,=\,8.779\,$\pm$\,0.005, $m_0$\,=\,10.11\,$\pm$\,0.03, $m_1$\,=\,0.56\,$\pm$\,0.01, $\gamma$\,=\,0.31\,$\pm$\,0.01 and $\beta$\,=\,2.1\,$\pm$\,0.4. Using the median stellar mass and star-formation rate in each bin, the gas-phase metallicity predicted by Equation \ref{eq:FMR}, agrees within 1\,--\,$\sigma$ of that calculated for the  \zra{0.6}{1.0} and \zra{1.2}{1.8} samples. This indicates the fundamental plane does not evolve with cosmic time and in high-redshift galaxies we are sampling this plane at higher star-formation rates, in agreement with other studies \citep[e.g.][]{Mannucci2011,Cresci2019, Sanders2020}

\changed{It has been proposed that the lower metallicity at a fixed stellar mass could also be explained by the lower metallicity of infalling gas at higher redshift and an increase in the mass loading factor of outflows and winds, that remove metal-enriched gas from the interstellar medium. Other studies suggest this evolution maybe driven by the redshift evolution of star formation efficiency \citep[e.g.][]{Lilly2013}, outflow strength \citep[e.g.][]{Lian2018,Sanders2020}, and gas fraction \citep[e.g.][]{Sanders2020}}. Lower than expected metallicities in high-redshift galaxies may also be related to the increased likelihood of galaxy interactions in the distant Universe \citep[e.g.][]{Bustamante2018}.

\subsection{The mass\,--\,metallicity plane}\label{KGES_metal_subsec:fzr}

\subsubsection{Fundamental properties and metallicity}

The correlations between the gas-phase metallicity of a galaxy and its fundamental properties are expected to go beyond a simple power-law correlation between stellar mass and metallicity. Both in the local Universe and at high-redshifts, studies have suggested the existence of a fundamental plane connecting gas-phase metallicity, stellar mass and star-formation rate. Galaxies of  higher star-formation rates are predicted to have lower metallicities at a given stellar mass, due to star-formation-driven winds and supernovae removing metal-enriched gas from the inter-stellar medium \changed{as well as the dilution by metal-poor gas accretion} \citep[e.g.][]{Ellison2008,Mannucci2010,Stott2013b,Cresci2019,Curti2020}.

In Figure \ref{KGES_metal_fig:NHratio_mstar}, we colour the galaxies by their dust-corrected H$\alpha$ star-formation rate as a first route to identify this plane in our sample. \changed{The \HA{} dust correction is derived from SED fitting, (see Section \ref{KGES_metal_sec:sample})}. The indication of a weak correlation is clear, with more highly star-forming galaxies being below the median trend between stellar mass and gas-phase metallicity. To more solidly establish the presence of the fundamental plane, we measure each galaxy's offset to the median stellar mass\,--\,metallicity  relation in Figure \ref{KGES_metal_fig:NHratio_mstar}. To do this we define the quantity $\Delta Z$, where $\Delta Z \equiv Z_{\rm gal}\,-\,Z_{\rm med}$. $Z_{\rm gal}$ is the gas-phase metallicity of the galaxy whilst $Z_{\rm med}$ is the gas-phase metallicity given by \changed{an interpolation of the} median track at the same stellar mass as the galaxy,  \changed{shown in Figure \ref{KGES_metal_sec:MZR}}. We use the median mass\,--\,metallicity relation for each redshift subsample sample to measure $\Delta Z$ for all 644 galaxies in the observational sample.

In Figure \ref{KGES_metal_fig:dz_corr}, we show $\Delta Z$ as a function of the H$\alpha$ derived star-formation rate, stellar continuum half-light radius (\rh{}), \HA{} specific star-formation rate (sSFR[yr$^{-1}$] $\equiv$ SFR$_{\rm H\alpha}/M_*$) and the balance between \HA{} rotational velocity and velocity dispersion (\vsigma)\footnote{The rotation velocity and velocity dispersion for each galaxy are measured at 2\rh{} from the galaxy's spatially-resolved \HA{} kinematics. See Tiley et al. in prep. for details.}. We identify a strong correlation between $\Delta Z$ and star-formation rate, with a Spearman coefficient of $r$\,=\,--\,0.79 and the probability the correlation is due to chance $p$\,=\,0.002. The linear parametric fit has a slope of $-$0.08\,$\pm$\,0.03 (i.e. 2.6\,--\,$\sigma$ from being flat), whereby more highly star-forming galaxies have lower gas-phase metallicities for a given stellar mass. A similar trend is identified individually in the \zra{0.6}{1.0} and \zra{1.2}{1.8} samples. To understand whether this correlation is driven by incompleteness at low star-formation rates and low metallicity, we exclude galaxies with $\rm \log_{10}(SFR [M_{\odot}\,yr^{-1}]) < 0.5$ and re-fit the data. We again find a strong negative correlation with $r$\,=\,--\,0.81  ($p$\,=\,0.002) at a 4.3\,--\,$\sigma$ level. We also fit to galaxies with $\log_{10}$(M$_*$)\,$<$\,10.0, to understand the impact of incompleteness at high stellar masses due to the removal of AGN. We find a weaker trend with  $r$\,=\,--\,0.54  ($p$\,=\,0.26) at a 1.2\,--\,$\sigma$ level, indicating this a potential driver of the correlation.

This is in agreement with other studies of gas-phase metallicity in star-forming galaxies at high redshift \cite[e.g.][]{Mannucci2010,Stott2013b,Zahid2014,Magdis2016,Sanders2018,Cresci2019}. The dependence on star-formation rate is expected to be driven by gas accretion and cold gas inflows, especially at high redshift, that leads to an increase in the star-formation rate and a dilution of the interstellar medium's metal content.

\begin{figure*}
\centering
\begin{minipage}{.5\textwidth}
  \centering
  \includegraphics[width=\linewidth,trim={0cm 1.5cm 0cm 0cm}]{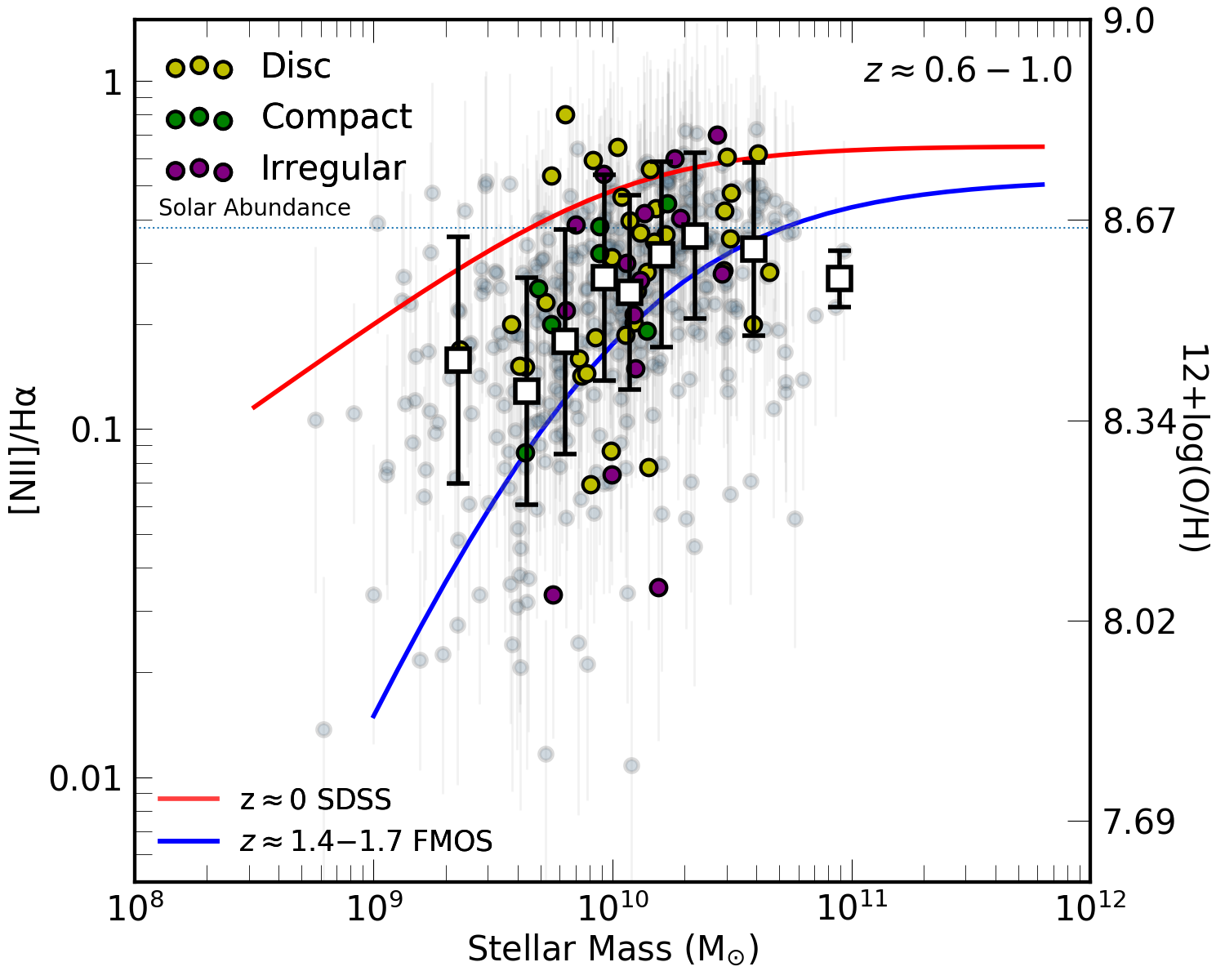}
\end{minipage}%
\begin{minipage}{.5\textwidth}
  \centering
  \includegraphics[width=\linewidth,trim={0cm 1.5cm 0cm 0cm}]{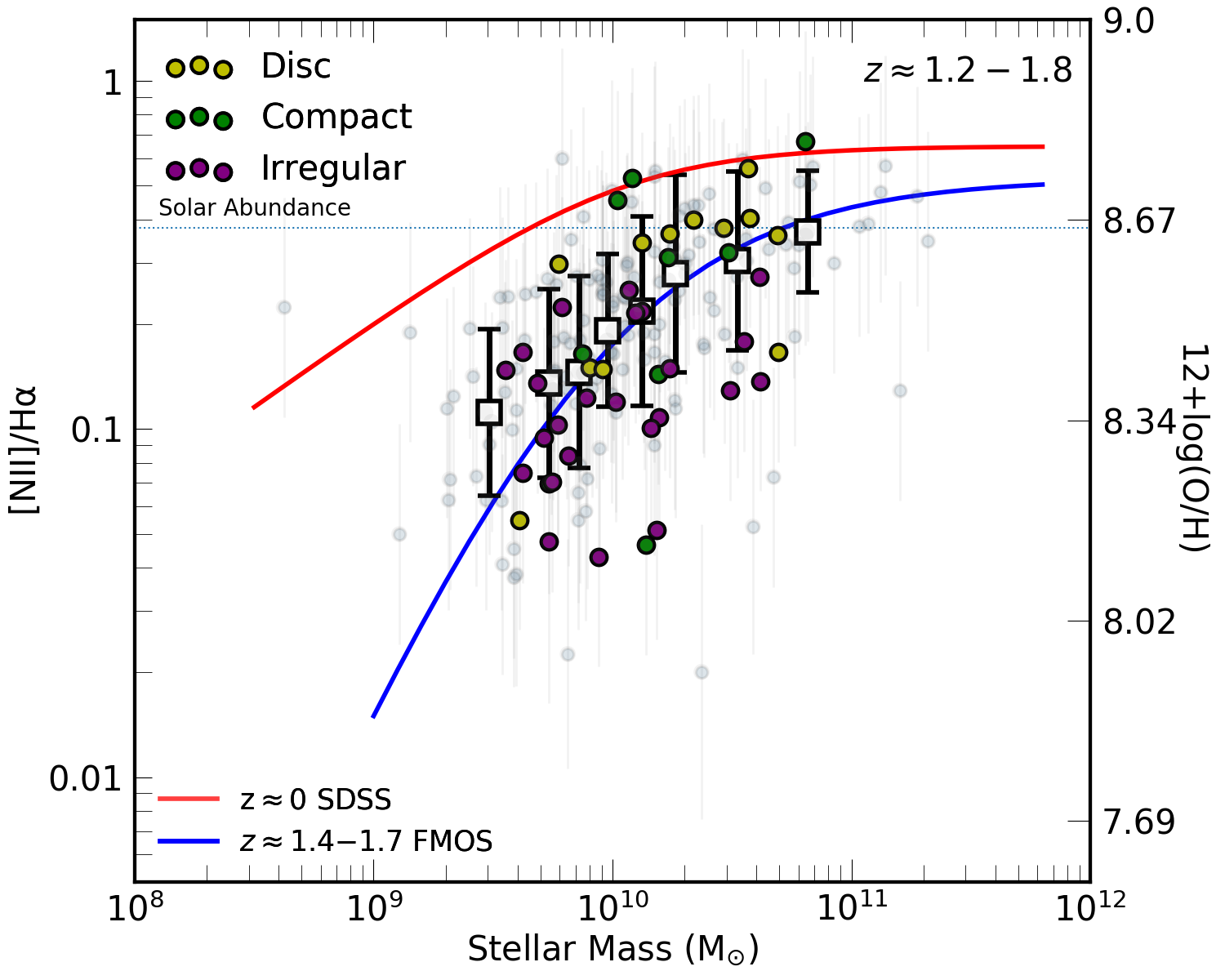}
\end{minipage}
\caption{Mass\,--\,metallicity relation of the \zra{0.6}{1.0} (\textbf{left}) and \zra{1.2}{1.8} (\textbf{right}) samples with data points coloured according to the \citet{Huertas-Company2015} morphological classification of disc, compact or irregular. Galaxies without a classification are shown by the grey points. The red line represents the mass\,--\,metallicity relation at $z$\,$\approx$\,0 derived from SDSS Survey \citep{Curti2020}, whilst the blue line indicates the mass\,--\,metallicity relation at \zra{1.4}{2.7} from the FMOS survey \citep{Zahid2014}. The horizontal dotted line indicates solar metallicity (12+log(O/H)\,=\,8.66; \citealt{Asplund2005}). The metallicity is converted from a \NH{} ratio to an Oxygen abundance following the calibration of \citet{Pettini2004}. The white squares (and error bars) represent running medians and 1\,--\,$\sigma$ scatters of the full sample. At \zra{0.6}{1.0} all morphological classes exhibit similar galaxy integrated gas-phase metallicity. At \zra{1.2}{1.8} irregular galaxies appear to have lower gas-phase metallicity than disc and spheroidal galaxies. Note the limit \NH{}$<$0.8 at \zra{0.6}{1.0} and \zra{1.2}{1.8} to remove AGN, which are predominantly higher stellar mass galaxies. }
\label{MZR_morphology}
\end{figure*}

As well as the ongoing star-formation rate of a galaxy correlating with the gas-phase metallicity, the specific star formation rate (sSFR) has been suggested to weakly anti-correlate with metallicity \citep[e.g.][]{Ellison2008,Mannucci2010,Pilyugin2013,Sanders2018,Huang2019}. For galaxies with $\log_{10}($sSFR[yr$^{-1}$])\,$\geq$\,$-$9.5, at a given stellar mass, higher specific star formation systems are predicted to have lower gas-phase metallicity, whilst galaxies with $\log_{10}($sSFR[yr$^{-1}$])\,$<$\,$-$9.5 have been shown to exhibit no significant correlation between sSFR and metallicity \cite[e.g.][]{Mannucci2010,Salim2015,Torrey2018}. For our sample, we identify a  moderate correlation between H$\alpha$ derived specific star-formation rate and  $\Delta Z$ with $r$\,=\,--\,0.56 ($p$\,=\,0.05) corresponding to  a 1.6\,--\,$\sigma$ significance which is in agreement with the anti-correlation identified by \citet{Magdis2016} between metallicity and specific star-formation rate main sequence offset ($\Delta$sSFR) in the KROSS survey and also the  3\--\,$\sigma$ negative correlation identified by \citet{Sanders2018} between gas-phase metallicity and specific star-formation rate in the MOSDEF survey of 260 star-forming galaxies  at $z$\,$\approx$\,2.3. Fitting to galaxies with $\log_{10}($sSFR[yr$^{-1}$])$\geq$\,$-$9.5, we identify a correlation of $r$\,=\,--\,0.75  ($p$\,=\,0.01) with a 2.5\,--\,$\sigma$ significance. 

Observational studies have also shown that the position of a galaxy in the mass\,--\,metallicity plane in the local Universe, and at high redshift, has a dependence on the physical extent of the galaxy \citep[e.g.][]{Ellison2008,Brisbin2012,Yabe2014,Huang2019}. Galaxies with larger stellar continuum half-light radii, for a given stellar mass, exhibit lower gas-phase metallicity. 
This correlation was also found by \citet{Sanchez2018} in the {\sc{eagle}} hydrodynamical simulation, with smaller star-forming galaxies predicted to have higher metallicity from $z \approx$\,0\,--\,8. To explore this correlation in our sample, in Figure \ref{KGES_metal_fig:dz_corr} we also analyse the correlation between $\Delta Z$ and a galaxy's stellar continuum half-light radius.

We identify a moderate negative correlation at 1.6\,--\,$\sigma$ between the continuum size of the galaxy and the offset from the median metallicity of the sample for a given stellar mass with a Spearman rank coefficient of $r$\,=\,--\,0.48  ($p$\,=\,0.12). Only $\approx$60 per cent of the sample have stellar continuum half-light radii derived from $HST$ imaging. If we exclude galaxies without $HST$ imaging, whose continuum sizes are less certain, we identify a correlation between $\Delta Z$ and continuum size of similar significance with $r$\,=\,--\,0.54  ($p$\,=\,0.22) at 2\,--\,$\sigma$.
This is in agreement with the correlation reported by other studies of star-forming galaxies \citep[e.g.][]{Ellison2008,Huang2019} and is attributed to smaller galaxies having higher stellar-mass surface densities and steeper potential wells, which are less effected by outflows and stellar winds removing metals from the interstellar medium.

Finally, in Figure \ref{KGES_metal_fig:dz_corr} we also show the correlation between $\Delta Z$ and \vsigma, identifying no correlation (Spearman rank coefficient of $r$\,=\,$-$0.19, $p$\,=\,0.55). This indicates that more rotation-dominated galaxies have similar gas-phase metallicity to more dispersion-dominated galaxies at a given stellar mass. \citet{Zenocratti2019} showed that in the {\sc{eagle}} simulation, more rotationally-supported galaxies at $z$\,=\,0 with $\log_{10}$($M_*$)\,$<$\,10.0 have lower metallicities, whilst at higher stellar masses more dispersion-dominated galaxies have lower metallicities. At higher redshift it was proposed that the correlation at lower stellar masses weakens whilst it is stronger for higher stellar-mass galaxies. 

Refitting the correlation to just galaxies with $\log_{10}$($M_*$)\,$>$\,10.0, i.e. galaxies in which nuclear activity is more prevalent and mergers play a key role, the parametric fit has a slope of 0.03\,$\pm$\,0.03 (i.e. consistent with no correlation) with a Spearman rank coefficient of $r$\,=\,0.43 ($p$\,=\,0.33). For galaxies with $\log_{10}$($M_*$)\,$<$\,10.0, i.e. galaxies in which star formation dominates the feedback, we find a moderate correlation with $r$\,=\,0.66 ($p$\,=\,0.15) and a parametric slope of 0.03\,$\pm$\,0.02 (i.e. 1.5\,--\,$\sigma$ significance). This is in contrast to the \citet{Zenocratti2019} results. It is well known that the rotation dominance of a galaxy is correlated with its morphology, with late-type disc galaxies exhibiting higher levels of rotation dominance, so next we explore the correlation between morphology and metallicity in our sample.

\subsubsection{Morphology and metallicity}

As well as the dynamical and photometric properties of a galaxy, a galaxy's morphology is known to correlate with its gas-phase metallicity \citep[e.g][]{Calura2009,Scudder2012,Gronnow2015,Horstman2020}. In the local Universe galaxies with disturbed morphologies have been shown to have metallicities different than those of isolated galaxies, and that depends on the magnitude of the galaxy-galaxy interactions \citep[e.g.][]{Michel-Dansac2008,Rupke2010,Bustamante2018}.

To explore this correlation within our sample, in Figure \ref{MZR_morphology} we show the mass\,--\,metallicity relation for the two galaxy samples, according to the galaxy rest-frame optical morphology as classified by \citet{Huertas-Company2015}. For \zra{0.6}{1.0} galaxies there is no variation of galaxy integrated gas-phase metallicity with morphological class.  At \zra{1.2}{1.8} irregular galaxies have on average 0.11\,$\pm$\,0.03\,dex lower metallicities than the running median of the sample  at a given stellar mass, compared to spheroidal or disc galaxies that have on average 0.04\,$\pm$\,0.03 and 0.06\,$\pm$\,0.03\,dex higher metallicities than the running median of the samples, respectively. To understand whether this trend is driven by sampling different regions of the galaxies between morphological classes, we also explore the correlation between the ratio of a galaxy's \HA{} extent to stellar extent (as parameterised by the stellar continuum half-light radius \rh) and morphological class, but we find no variation between spheroidal, disc and irregular galaxies.

In \citet{Gillman2019b}, irregular galaxies at \zra{1.2}{1.8} were inferred to have higher gas fractions in comparison to disc galaxies. A higher fraction of metal poor gas, potentially caused by inflows of low-metallicity gas due to the tidal forces and gravitational torques associated with previous galaxy interactions \citep[e.g][]{Bustamante2018}, dilutes the metal content and leads to a lower observed metallicity at a given stellar mass, as shown in Figure \ref{MZR_morphology}. 
The irregular galaxies at \zra{1.2}{1.8} have a median stellar mass of \med{$\log$($M_*$[$M_{\odot}$])}=\,9.95 whilst the disc and spheroidal galaxies have \med{$\log$($M_*$[$M_{\odot}$])}=\,10.28 and \med{$\log$($M_*$[$M_{\odot}$])}=\,10.13, respectively. The median H$\alpha$ star-formation rate of the irregular galaxies at \zra{1.2}{1.8} is \med{SFR}=\,24\,$\pm$\,5\,\SFR whilst disc and spheroidal galaxies have \med{SFR}=\,26\,$\pm$\,11 and \med{SFR}=\,32\,$\pm$\,10 \SFR respectively. 

Using the parameterisation of the fundamental plane between stellar mass, star-formation rate and gas-phase metallicity given by \citet{Curti2020}, we derive the expected metallicity of disc, spheroidal and irregular galaxies, if all three samples lie on the fundamental plane. Using Equation \ref{eq:FMR}, we identify that disc and spheroidal galaxies at \zra{1.2}{1.8} have a median fundamental plane (FMR) metallicity of \med{$Z_{\rm disc, FMR}$}\,=\,8.59\,$\pm$\,0.02 and  \med{$Z_{\rm spheroidal, FMR}$}\,=\,8.52\,$\pm$\,0.01, respectively. Irregular galaxies on the fundamental plane have a median metallicity of \med{$Z_{\rm irregular,FMR}$}\,=\,8.50\,$\pm$\,0.01.

As we show in Figure \ref{MZR_morphology}, the disc and spheroidal galaxies of our sample at \zra{1.2}{1.8} have an actual median metallicity of \med{Z$_{\rm disc}$}\,=\,8.64\,$\pm$\,0.06 and \med{Z$_{\rm spheroidal}$}\,=\,8.61\,$\pm$\,0.11 respectively. Whilst irregular galaxies have a median metallicity of \med{Z$_{\rm irregular}$}\,=\,8.39\,$\pm$\,0.03, in comparison to the solar abundance $Z_{\odot}$\,=\,8.66 measured by \citet{Asplund2005}. This indicates that the irregular galaxies identified in our \zra{1.2}{1.8} sample do not lie on the fundamental mass\,--\,metallicity  relation established by \citet{Curti2020}, in contrast to disc and spheroidal galaxies, that do lie on or near the plane. A similar result was also obtained by \citet{Bustamante2020} in the SDSS survey at $z$\,$\approx$\,0. \citet{Bustamante2020} demonstrated that galaxies showing signs of interactions lie off the fundamental plane,  exhibiting lower gas-phase metallicities. They concluded that the offset from the fundamental plane is driven by the recent triggering of gas inflows, that are stronger than those normally experienced by the `main-sequence' star-forming galaxies that the fundamental plane describes.

\section{Metallicity Gradients}\label{KGES_metal_Sec:zgrad}

Constraining the distribution of gas-phase metallicity in a galaxy provides insights into the baryonic processes (e.g. star formation, feedback and accretion) that dominate a galaxy's evolution, and ultimately lead to changes in galaxy dynamics and morphology. Utilizing the spatially-resolved dynamics of the galaxies in our sample, we can analyse how a galaxy's gas-phase metallicity is correlated with radius and how this is connected to other galaxy properties such as those presented in Figure \ref{KGES_metal_fig:dz_corr} (e.g. \HA{} star-formation rate and stellar continuum size).

In this section we analyse the \NH{} line ratio (i.e metallicity) gradients in our galaxies, including modelling of the systematics encountered in our analysis and consistency tests, in addition to correlating the metallicity gradients of the observational sample with the morphological and dynamical properties of the galaxies.

\subsection{Radial metallicity profiles}

To measure the radial dependence of the gas-phase metallicity of each, we sum the spectra in the de-redshifted  data cube in elliptical annuli whose semi-major axes are multiples of the half-light radius of the \changed{PSF of the integral field data}. The axis ratio, position angle and half-light radius of the galaxy are derived from $HST$ CANDELS images (see \citealt{Gillman2019b}). We fit a three Gaussian profiles to the H$\alpha$ and [{N\sc{ii}}] emission lines in each annulus as before, from which we measure the \NH{} ratio. To fit the emission lines we require the integrated H$\alpha$ line of the annulus to have a signal-to-noise (S/N)\,$\geq$\,3.  We extract the \NH{} ratio in all annuli where these criteria are met. If the [{N\sc{ii}}] emission line has a S/N\,$\leq$\,3 we instead calculate a 3\,--\,$\sigma$ upper limit for \NH{} ratio (see Figure \ref{KGES_metal_fig:NHratio_spec} for examples).

The emission lines are fit using the same $\rm \chi^2$ minimisation procedure as for the integrated spectra, with the FWHM and wavelength separation of the two [{N\sc{ii}}] lines and H$\alpha$ line fixed. Examples of the \NH{} ratio measurements for a number of galaxies are shown in Appendix \ref{App:profile}. If we have more than three measurements of \NH{} as a function of radius with at least one measurement not being a limit, we additionally measure the slope of the \NH{} ratio as a function of radius.

\begin{figure*}
\centering
\includegraphics[width=\textwidth,trim={0cm 2cm 0cm 0cm}]{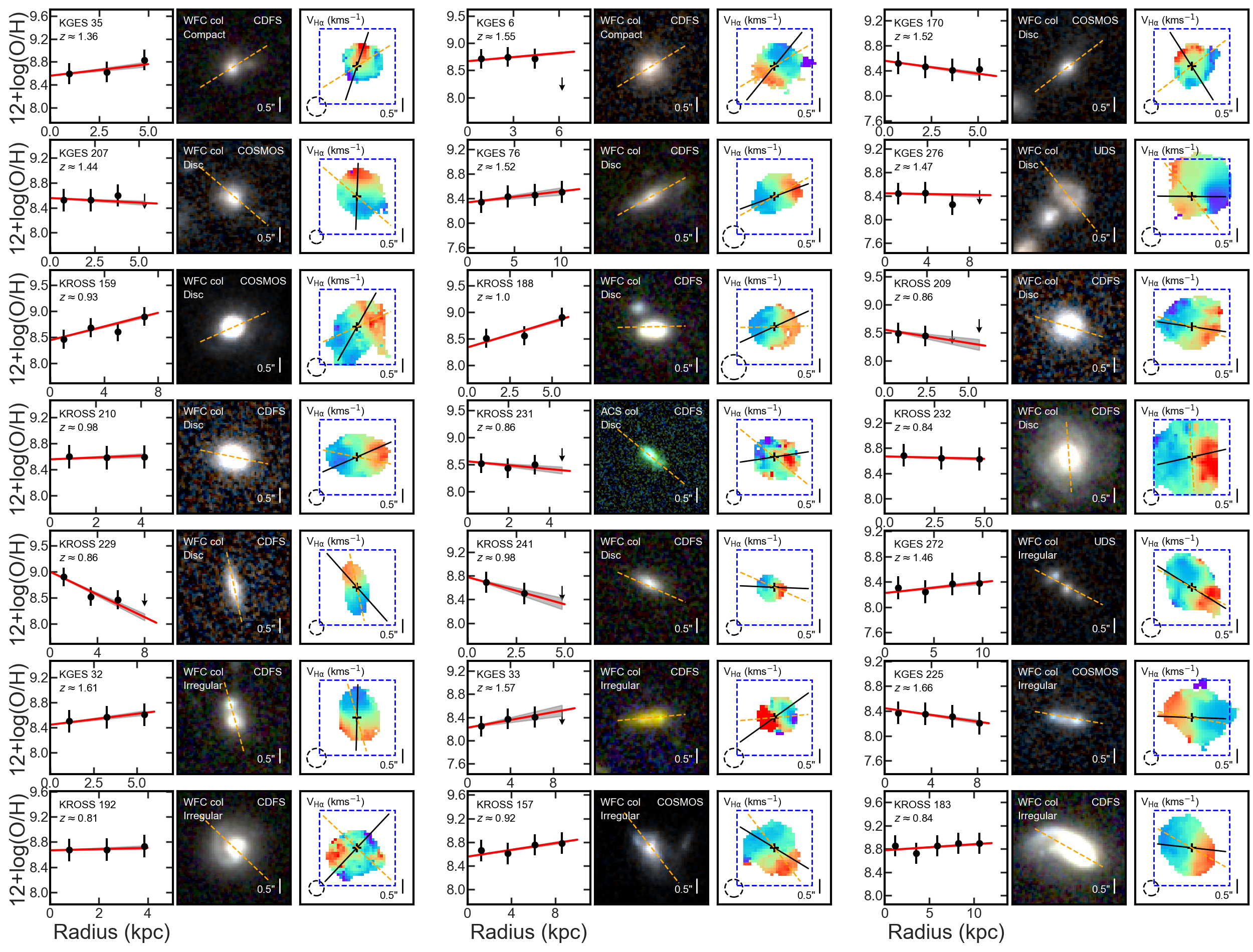}
 \caption{Examples of spatially-resolved compact, disc and irregular galaxies with measured \NH{} gradients. For each galaxy, we show in the left panel the radial metallicity profile (black data points) and linear parametric fit (red solid line) with 1\,--\,$\sigma$ error (grey shaded region), and list the galaxy name and the \HA-derived $z_{\rm spec}$ in the top-left corner. A $HST$ colour image with the morphological position angle overlaid (orange dashed line) is shown in the central panel, where we also list the $HST$ instrument used, galaxy type and extragalactic deep field. Finally, we show in the right panel, the H$\alpha$ velocity map with kinematic position angle (black solid line), kinematic centre (black cross), the FWHM of the seeing (black dashed circle) and KMOS IFU field of view (blue dashed square) overlaid. There is no correlation between the metallicity gradients of the galaxies and their rest-frame optical morphologies classified from the $HST$ images.}
  \label{KGES_metal_fig:MHgrad_kin} 
\end{figure*}

\subsection{Consistencies and calibrations}

Since our observations are ground-based and are sensitive to variations in galaxy properties (such as galaxy size or inclination) as well as seeing (see Section \ref{KGES_metal_sec:BS}), it is important to quantify the reliability of our method to derive the intrinsic metallicity gradient of a galaxy.
To test the reliability of the derived metallicity gradient, we analyse the effects of extracting the spectra in annuli with different position angles (e.g. PA$_{\rm kin}$, PA$_{\rm morph}$) and with varying semi-major axes (e.g. 2\,kpc, 0.5\rh, $R_{\rm h,psf}$). The results are presented in \citet{Gillman_thesis}, where we compared the metallicity gradients derived with different position angles and semi-major axes. \changed{We also explore the correlation between metallicity gradient and the axis ratio of the galaxy, identifying no correlation.}

We demonstrated in \citet{Gillman_thesis} that the choice of position angle and annulus size has no systematic impact on the measured \NH{} metallicity gradients. We thus conclude that our method of using annulus sizes relative to the size of the \changed{PSF of the integral field data, aligned} with the galaxy's kinematic position angle, and adopting 3\,--\,$\sigma$ upper limits for low signal-to-noise regions to measure the metallicity gradients is robust and does not bias the measured gradients \changed{and uncertainties}.

\subsection{Beam smearing}\label{KGES_metal_sec:BS}

One of the most significant systematic effects that affects our ability to accurately measure the metallicity gradients of high-redshift galaxies is beam smearing \citep[e.g.][]{Yuan2013,Mast2014,Carton2017,Acharyya2020}. The atmospheric seeing blurs the observed emission of each galaxy. The metallicity gradient measured is thus expected to be artificially flattened compared to the intrinsic gradient of the galaxy. 

To quantify this effect and understand the impact on our measurements we generate 1000 mock galaxies with a given intrinsic metallicity gradient ranging from $-$0.07 to 0.07\,dex\,kpc$^{-1}$. We then build mock datacubes with the same properties as our KMOS observations, convolve each velocity slice with the PSF and measure the observed metallicity gradient of the galaxy. 

For these simulations, we model the H$\alpha$ spatial distribution of the galaxies using two dimensional S\'ersic profiles with an index of $n$\,=\,1 (exponential disc) and a range of half-light radii, ellipticity and position angles that match the distribution of these observables in the  \zra{1.2}{1.8} and  \zra{0.6}{1.0} samples. We assume the H$\alpha$ half-light radius is equal to the stellar continuum half-light radius of each galaxy, and use this to define the annulus sizes from which we extract the \NH{} ratios. To model the PSFs of the observations we use a circular two-dimensional Gaussian function, that is scaled relative to the size of the model galaxy. The velocity profiles of the input emission lines are a combination of Gaussian profiles and Gaussian noise, scaled to match the required signal-to-noise.

The results of the beam smearing modelling analysis is discussed in Appendix \ref{App:BS}.  We derive a beam-smearing correction factor that is a function of FWHM/\rh{} and the axis ratio. We note we do not account for a dependence on the intrinsic metallicity gradient of the galaxy. As expected, galaxies with a larger FWHM/\rh{} ratio i.e a smaller stellar continuum half-light radius (\rh) at fixed PSF FWHM, are more affected by beam-smearing. In addition at fixed FWHM/\rh, more edge-on galaxies also have a larger correction factor. To correct the metallicity gradients measured in our observational sample for beam smearing, we interpolate the observed-to-intrinsic gradient ratios as a function of FWHM/R$_{\rm h}$ for a given axis ratios as shown in Appendix \ref{App:BS}. We use the resulting correction curve to derive the beam smearing correction factor of each galaxy in our sample. \changed{The median beam smearing correction factor for the  \zra{0.6}{1.0} sample is $C_{\rm BS} $\,=\,0.66\,$\pm$\,0.17 whilst for the \zra{1.2}{1.8} galaxies we derive a median factor of $ C_{ \rm BS}$\,=\,0.64\,$\pm$\,0.18}.

For galaxies with FWHM/\rh{} ratios above $\approx$1, the observed metallicity gradient is $<50$ per cent of the intrinsic metallicity gradient and is therefore very uncertain. For the remainder of our analysis, we thus limit the sample and present the beam smearing-corrected metallicity gradients only in galaxies with a correction factor $<$50 per cent (\changed{98} galaxies). This sub-sample of galaxies has a distribution of stellar mass similar to that of the full observational sample whilst exhibiting larger stellar continuum sizes (see Appendix \ref{App:BS}). \changed{We note sub-sample shows a weak negative correlation between metallicity gradient and axis ratio that is not present in the full sample. This is due to incompleteness at low axis ratio values, as these galaxies have significant beam smearing correction factors.}

The median beam smearing-corrected metallicity gradient, for the sample of \changed{98} galaxies with robust beam smearing-corrected metallicity gradients, is \changed{$\Delta Z / \Delta R$\,=\,0.002\,$\pm$\,0.004\,dex\,kpc$^{-1}$} (i.e. consistent with no gradient) with a 16\,--\,84th percentile range of \changed{$\Delta Z/ \Delta R$\,=\,$-$0.034\,--\,0.043\,dex\,kpc$^{-1}$.} We now explore the correlations between the galaxies beam smearing-corrected metallicity gradients and their morphological and dynamical properties. 

\subsection{Metallicity gradients and morphology}

In the local Universe studies of the metallicity gradients in star-forming galaxies have established a link between galaxy morphology and the slope of the radial metallicity profile. Interacting galaxies have been shown to exhibit flattened gradients \citep[e.g.][]{Kewley2010,Rupke2010,Taylor2017}, which is anticipated from inflows of low-metallicity gas into the galaxy centres as a result of the interactions. At high redshift integral-field studies of star-forming galaxies with irregular morphologies have revealed them to have inverted (positive) metallicity gradients, i.e metallicity increasing with radius \citep{Queryel2012}.

Early- and late-type galaxies locally have been shown to exhibit a range of radial metallicity gradients, with some studies identifying no morphological dependence \citep[e.g.][]{Zaritsky1994,Sanchez2014,Sanchez-M2016} whilst others indicate that early-type galaxies have no gradient but late-type galaxies have seep (negative) abundance profiles \citep[e.g.][]{Marquez2002}.

In Figure \ref{KGES_metal_fig:MHgrad_kin} we show examples of the kinematics and metallicity gradients of galaxies in our sample with $HST$ imaging, for a range of rest-frame optical morphologies as determined by \cite{Huertas-Company2015}. There is no correlation between the rest-frame optical morphologies of the galaxies, as indicated by the $HST$ images, and the gradients of the metallicity profiles, with disc, spheroidal and irregular galaxies having a range of metallicity gradients.
To search for a potential link between the gas-phase metallicity gradient and the rest-frame optical morphology of a galaxy, in Table \ref{tab:Zgrad} we show the median beam smearing-corrected gas-phase metallicity gradients of spheroidal, disc and irregular galaxies at \zra{0.6}{1.0} and \zra{1.2}{1.8}. At \zra{0.6}{1.0} there are no spheroidal (compact) galaxies and at \zra{1.2}{1.8} there is only one spheroidal galaxy with a beam smearing correction factor less than 50 per cent. There is no difference between the metallicity gradients of the three morphological classes, with all galaxies having a median metallicity gradient within 1\,--\,$\sigma$ of each other, even though their integrated gas-phase metallicities are different at \zra{1.2}{1.8} (Figure \ref{MZR_morphology}). 

\begin{table}
    \centering
    \caption{Median beam smearing-corrected gas-phase metallicity gradients in spheroidal, disc and irregular galaxies at \zra{0.6}{1.0} and \zra{1.2}{1.8}.}
    \begin{tabular}{l|c|c|}
    \hline
     \multirow{ 2}{*}{Morphology} & \multicolumn{2}{|c|}{$\Delta Z$/$ \Delta R$ [dex kpc$^{-1}$]} \\
     \cline{2-3}
      & \zra{0.6}{1.0}& \zra{1.2}{1.8}\\
     \hline
    
      Compact  & --                          & \changed{\phantom{$-$}0.037\,$\pm$\,0.004}     \\
     Disc      & \changed{$-$0.003\,$\pm$\,0.010 }   & \changed{\phantom{$-$}0.005\,$\pm$\,0.011}    \\
    Irregular   & \changed{\phantom{$-$}0.011\,$\pm$\,0.023 }      & \changed{$-$0.006\,$\pm$\,0.017} \\
     
     \hline
    \end{tabular}
    \label{tab:Zgrad}
\end{table}

\begin{figure*}
    \centering
    \includegraphics[width=\linewidth,trim={0cm 2cm 0cm 0cm}]{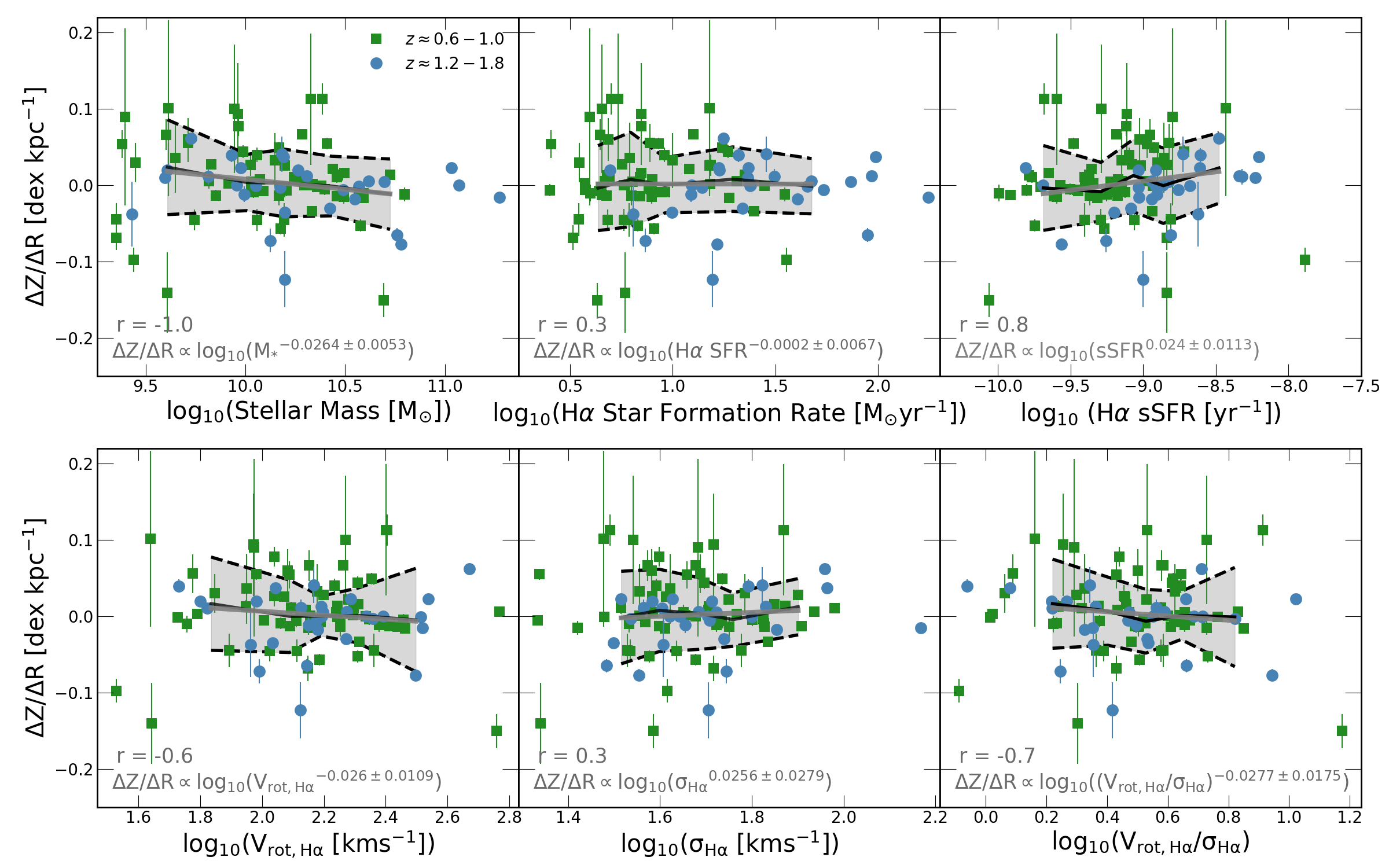}
    \caption[Metallicity gradients correlations]{Bean smearing-corrected metallicity gradients of the galaxies our sample as a function of their stellar masses, dust-corrected H$\alpha$ derived star-formation rates, H$\alpha$ specific star-formation rates, rotational velocities, velocity dispersions and ratio of rotational velocity to velocity dispersion. The black dashed lines and grey-shaded regions indicate the 1\,--\,$\sigma$ scatters of the running medians (black solid lines). We also show a linear parametric fit to the running median of each relation, as shown in the lower-left corner of each panel and quantified by the Spearman rank coefficient.  We identify a strong negative correlation with the stellar masses and a moderate positive correlation with specific star-formation rates but no significant correlation with the star-formation rates of the galaxies. The metallicity gradients of the galaxies show an insignificant positive correlation with the velocity dispersions, but a moderate positive (negative) correlation with the rotational velocities (rotation dominance, \vsigma{}).}
    \label{KGES_metal_fig:MHgrad_corr}
\end{figure*}

Recent studies of high-redshift star-forming galaxies have also established there is no correlation between a galaxy's metallicity gradient and its morphology. \citet{Curti2019} identified no difference between the metallicity gradients of interacting and disturbed galaxies compared to non-interacting systems in a sample of 42 lensed galaxies at \zra{1.2}{2.5}. They attributed this to the limited spatial resolution of their high-redshift data as well as the method of averaging the metallicity in annuli and thus smoothing out azimuthal variations of metallicity. High-resolution studies of local star-forming galaxies have also demonstrated the importance of resolving individual H{\sc{ii}} regions and azimuthal variations when analysing the connection between morphology and metallicity profiles \citep[e.g.][]{Sanchez2017b,Sanchez-M2018}.

\subsection{Metallicity gradients and fundamental properties}

At low redshifts strong trends between the stellar masses, specific star-formation rates and metallicity gradients of star-forming galaxies have been observed \citep[e.g.][]{Sanchez2014,Belfiore2017,Belfiore2019}. Less massive, high specific star-formation rate systems exhibit flatter metallicity gradients, whilst more massive, low specific star-formation rate galaxies have steeper (more negative) abundance gradients.

To analyse whether the metallicity gradients of the galaxies in our sample are correlated to other observable properties besides morphology, in Figure \ref{KGES_metal_fig:MHgrad_corr} we plot their metallicity gradients as a function of stellar mass, dust-corrected H$\alpha$ derived star-formation rate and  H$\alpha$ specific star-formation rate.
The stellar masses of the galaxies show a strong negative correlation with the metallicity gradients, with a Spearman rank coefficient of \changed{$r$\,=\,$-$1.0 ($p$\,=\,0) and a slope of $-$0.03\,$\pm$\,0.005 (5\,--\,$\sigma$ from flat)}. This indicates that higher stellar mass galaxies in our sample have steeper (more negative) metallicity gradients. The H$\alpha$ star-formation rates indicates a no trend with the metallicity gradients with  \changed{$r$\,=\,0.3  ($p$\,=\,0.62) and a non-zero slope at a  0.2\,--\,$\sigma$} significance. The specific star-formation rate of the galaxies shows a moderate positive trend with  \changed{$r$\,=\,0.80 ($p$\,=\,0.1) and a non-zero slope at a 2\,--\,$\sigma$} level. Galaxies with a lower specific star-formation rates in our sample have steeper (more negative) metallicity gradients.

\begin{figure*}
    \centering
    \includegraphics[width=0.9\linewidth,trim={0cm 1cm 0cm 0cm}]{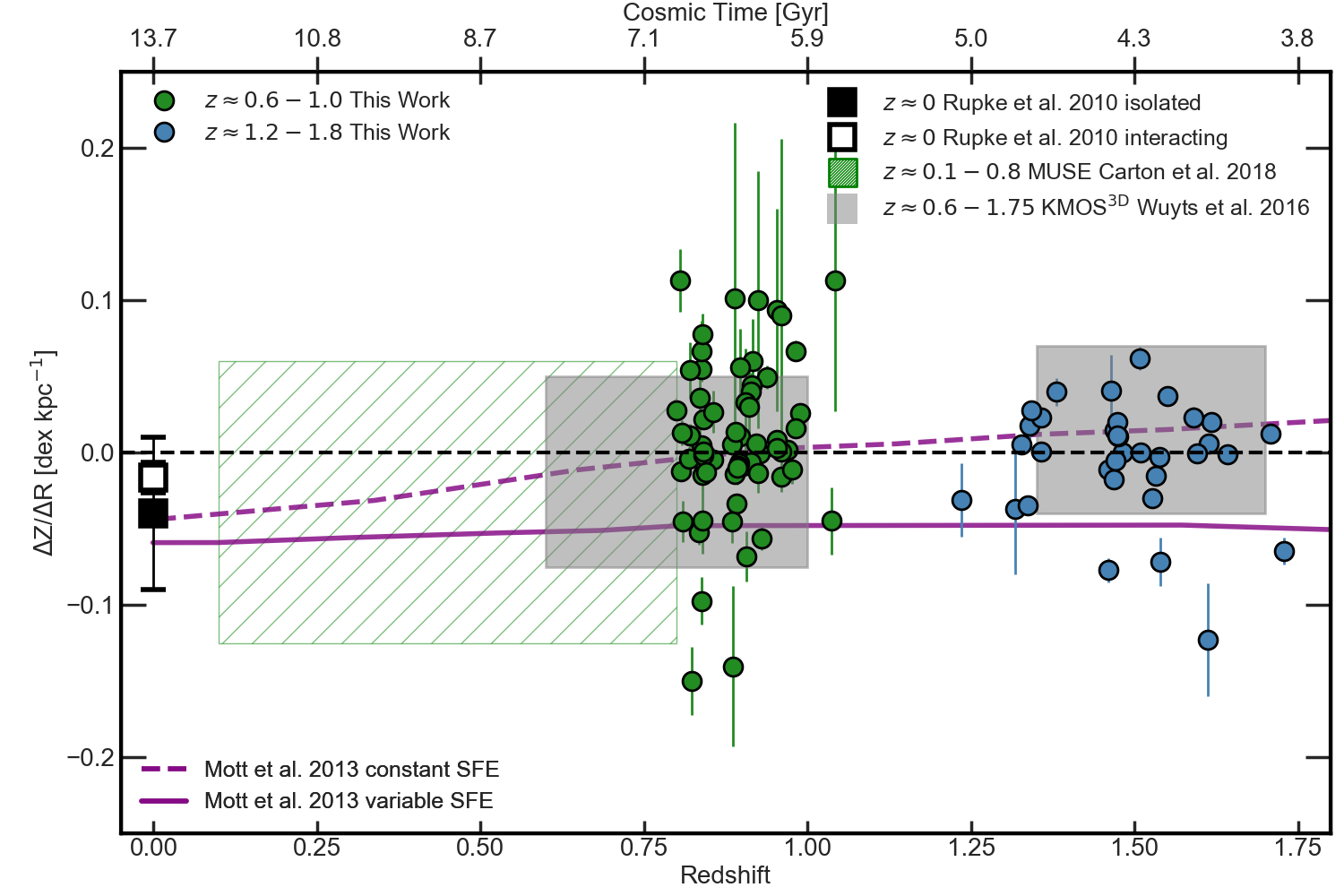}
    \caption[Cosmic evolution of metallicity gradients]{Beam smearing-corrected metallicity gradients of the galaxies in our sample as a function of their redshifts. We also show the metallicity gradients of a $z$\,$\approx$\,0 sample of isolated and interacting star-forming galaxies from \citet{Rupke2010} (respectively black and white square with error bars). The green hashed region indicates measurements from MUSE observations of intermediate-redshift galaxies from \citet{Carton2018}. The metallicity gradients derived by \citet{Wuyts2016} from the KMOS$^{\rm 3D}$ survey are indicated by the grey shaded regions. We also show theoretical predictions from two models of the metallicity gradients of disc galaxies from \citet{Mott2013} with radially constant star-formation efficiency (purple dashed line) and \changed{variable} star formation efficiencies (solid purple line). The metallicity gradients measured from our sample are in agreement with other observations of star-forming galaxies at high redshifts as well as theoretical models. There is no significant evolution with redshift between the redshift samples with both samples exhibiting flatter gradients than observed locally.}
    \label{KGES_metal_fig:MHgrad_z}
\end{figure*}

The correlation between specific star formation rates and radial metallicity gradients was also identified by other studies of high-redshift star-forming galaxies \citep[e.g.][]{Stott2014,Wuyts2016,Curti2019}, whereby galaxies with higher specific star-formation rates have metal-poorer centres. Hydrodynamical simulations have highlighted the importance of feedback in driving this correlation and shaping the metal distribution within galaxies \citep[e.g.][]{Ma2017}, suggesting that in these lower stellar mass, higher specific star-formation rate galaxies, feedback is more efficient. The negative correlation between metallicity gradient and stellar mass is also consistent with with the inside-out model of galaxy evolution \citep[e.g.][]{Dave2011,Gibson2013}, whereby the inner regions of galaxies form stars at earlier times, leading to an increase in the metallicity in the central regions as the galaxies evolves. 

Models of inside-out galaxy evolution also predict inflows of low-metallicity gas into the central regions of galaxies at early times, along filaments, that dilute the local metal distributions and boost the specific star-formation rate in the central regions. This may well be reflected in the tentative correlation seen in Figure \ref{KGES_metal_fig:MHgrad_corr}  between the metallicity gradients and H$\alpha$ specific star-formation rates of the galaxies in our sample.

Figure \ref{KGES_metal_fig:MHgrad_corr} also shows the correlation between the metallicity gradients of galaxies in the observed sample and their rotation velocities, velocity dispersions and the ratio of rotation velocity and velocity dispersion. We establish a moderate 2.3\,--\,$\sigma$ correlation between the metallicity gradient of a galaxy and its rotation velocity with a correlation coefficient of \changed{$r$\,=\,$-$0.60 ($p$\,=\,0.28)}. Whilst we find no correlation with the velocity dispersion \changed{($r$\,=\,0.30, $p$\,=\,0.62)}.

In the local Universe, studies of late-type galaxies in the MaNGA survey \citep[e.g.][]{Pilyugin2019} have shown that the metallicity gradients flatten with increasing rotation velocity, in contrast to our trend in Figure \ref{KGES_metal_fig:MHgrad_corr}. Similarly \citet{Queryel2012} established that, for star-forming galaxies at $z$\,$\approx$\,1.2 in the Mass Assembly Survey with SINFONI in VVDS (MASSIV), galaxies with higher gas velocity dispersions have shallower or more positive metallicity gradients in agreement with Figure \ref{KGES_metal_fig:MHgrad_corr}. We however identify a moderate negative correlation between the ratio of rotation velocity to velocity dispersion (\vsigma) of a galaxy and its metallicity gradient, with  \changed{$r$\,=$-$0.70 ($p$\,=\,0.18) and a non-zero slope at a 1.6\,--\,$\sigma$ level}. 

Other studies of high-redshift galaxies have also identified that more dispersion dominated galaxies have flatter metallicity gradients \citep[e.g.][]{Jones2013,Leethochawalit2016,Ma2017,Hemler2020}. It is expected that this correlation is driven by metal-rich gas being more effectively redistributed in dispersion dominated galaxies, thus flattening the metallicity gradients. 

\subsection{Cosmic evolution of metallicity gradients}\label{KMOS_metal_sec:zevo}

Finally we compare the metallicity gradients of the galaxies in our sample to other observational studies of the chemical abundance gradients of star-forming galaxies across cosmic time. We also compare our results to cosmological hydrodynamical simulations which trace the gas-phase metallicities of disc galaxies from $z$\,$\approx$\,2 to the present day.

In Figure \ref{KGES_metal_fig:MHgrad_z} we show the metallicity gradients derived by \citet{Rupke2010} at $z$\,$\approx$\,0 for a sample of isolated spiral galaxies as well as galaxies undergoing interactions with other systems. The interacting systems have slightly flatter gradients, with,  $\langle$\,$\Delta Z$/$\Delta R$\,$\rangle$\,=\,0.02\,$\pm$\,0.05  dex kpc$^{-1}$  in comparison to isolated systems with $\langle$\,$ \Delta Z$/$ \Delta R $\,$\rangle$\,=\,$-$0.04\,$\pm$\,0.05 dex kpc$^{-1}$. 

At intermediate redshift we use the distribution of metallicity gradients of star-forming galaxies from \zra{0.1}{0.8} derived by \citet{Carton2018} using Multi Unit Spectroscopic Explorer (MUSE) observations. These metallicity gradients are derived using a combination of strong forbidden lines and Balmer emissions lines. In Figure \ref{KGES_metal_fig:MHgrad_z} we also show the \NH{} metallicity gradients derived by \citet{Wuyts2016} for the KMOS$^{\rm 3D}$ survey at \zra{0.6}{1.8}, which are largely in agreement with those measured for our sample.

Finally, to compare to the predictions of hydrodynamical simulations, we show the tracks from \citet{Mott2013} who model the evolution of abundance gradients in spiral galaxies in the framework of inside-out disc formation. We show two models, one with variable star-formation efficiencies within galaxy discs, and one with a fixed star-formation efficiency that recreates the inverted gradients seen at high redshifts. \citet{Mott2013} note that the inversion of the gradients is predominantly driven by efficient feedback mechanisms and the infall of pristine gas, counteracting chemical enrichment, at early times.

Figure \ref{KGES_metal_fig:MHgrad_z} indicates no significant evolution of the metallicity gradients with redshift between the \zra{0.6}{1.0} and \zra{1.2}{1.8} samples. At \zra{0.6}{1.0} the median metallicity gradient is \changed{ $\Delta Z/\Delta R$\,=\,$-$0.004\,$\pm$\,0.005} dex kpc$^{-1}$ with a 16\,--\,84th percentile range of \changed{ $-$0.03\,--\,0.05} dex kpc$^{-1}$. At \zra{1.2}{1.8} the median metallicity gradient is \changed{ $ \Delta Z/ \Delta R $\,=\,0.0002\,$\pm$\,0.0048 dex kpc$^{-1}$} with a 16\,--\,84th percentile range of \changed{$-$0.03\,--\,0.02 dex kpc$^{-1}$.} 

Our sample exhibits flatter metallicity gradients than those observed locally but is consistent with previous studies of the metallicity gradients of high-redshift star--forming galaxies \citep[e.g.][]{Wuyts2016,Carton2018,Curti2019}. This flattening of metallicity gradients at earlier cosmic times (higher redshift) is also found in hydrodynamical simulations when strong feedback processes are implemented \citep[e.g][]{Mott2013,Tissera2019,Hemler2020}. 

The  flattening of a galaxy's metallicity gradient could be caused \changed{by} a number of processes. Galaxy interactions and mergers can lead to radial inflows of metal-poor gas, resulting in the dilution of the central metal abundance \citep[e.g.][]{Rupke2010}. Feedback processes such star-formation-driven winds and supernovae, which are ubiquitous at high redshifts also act to reduce the metallicities of the central regions of star-forming galaxies in inside-out models of galaxy evolution \citep[e.g.][]{Newman2012,Swinbank2019,Schreiber2019}.

\section{Conclusions}\label{KGES_metal_Sec:Conc}

We have analysed the gas-phase metallicities of 644 typical star-forming galaxies at \zra{0.6}{1.8}, utilising the \NH{} emission line ratio as a probe. Using the spatially-resolved H$\alpha$ kinematics and morphological properties of the galaxies we, analyse the connection between gas-phase metallicity, galaxy dynamics and morphology. We summarize our findings as follows:
\begin{itemize}
    \item We establish that a gas-phase metallicity\,--\,stellar mass relation is present at $z$\,$\approx $\,0.8 and $z$\,$\approx$\,1.5 (Figure \ref{KGES_metal_fig:NHratio_mstar}), with more massive galaxies exhibiting higher metallicity. We demonstrate that the \zra{0.6}{1.0} sample galaxies with a median stellar mass of $  \log(M_*[M_{\odot}])$\,=\,10.0 has a median gas-phase metallicity of $\rm 12+\log(O/H)$\,=\,8.56\,$\pm$\,0.01. Whilst at \zra{1.2}{1.8}, the galaxies in our sample with a median stellar mass of  $ \log(M_*[M_{\odot}])$\,=\,10.0 have a median gas-phase metallicity of $\rm 12+\log(O/H) $\,=\,8.49\,$\pm$\,0.02.
    
    \item At a fixed stellar mass, galaxies with  higher \HA-derived star-formation rates (and specific star-formation rates) have lower gas-phase metallicity indicating the presence of a mass\,--\,metallicity\,--\,star-formation rate fundamental plane (Figure \ref{KGES_metal_fig:dz_corr}) with $\Delta Z \propto \log_{10}(SFR^{-0.08\pm0.03})$. We also identify that at fixed stellar mass, smaller galaxies (higher surface mass density) have higher gas-phase metallicities, with $\Delta Z \propto \log_{10}(R_{\rm h}^{-0.1\pm0.06})$. We establish there is no dependence on the rotation dominance of the galaxy and its position in the mass\,--\,metallicity plane.
    
    \item Utilizing the morphological classifications of  \citet{Huertas-Company2015}, we determine that at \zra{0.6}{1.0} spheroidal, disc and irregular galaxies exhibit similar gas-phase metallicities whilst at \zra{1.2}{1.8}, irregular galaxies have 0.11\,$\pm$\,0.03\,dex lower metallicities on average than disc and spheroidal galaxies, which are on average 0.04\,$\pm$\,0.03 and 0.06\,$\pm$\,0.03\,dex above the median metallicity of the sample galaxies (Figure \ref{MZR_morphology}). We attribute these lower metallicities to the higher gas fractions found in irregular galaxies at $z \approx 1.5$ \citep{Gillman2019b}.
    
    \item  To quantify the metallicity gradients of our sample we measure the \NH{} emission line ratio as a function of galactocentric radius using the spatially-resolved \HA{} data. We explore the correlation between metallicity gradient and other galaxy properties as well as the impact of beam smearing  on our measurements of the metallicity gradient. 
    For the sub-sample of  \changed{98} galaxies for which our simulations suggest a reliable intrinsic metallicity gradient can be inferred (median $\log(M_*[M_{\odot}]) $\,=\,10.16 and SFR\,=\,8\,$\pm$\,1\SFR{}), the median intrinsic metallicity gradient is \changed{$ \Delta Z$/$ \rm \Delta R$\,\,=\,0.002\,$\pm$\,0.004 dex kpc$^{-1}$  with a 16\,--\,84th percentile range of $\Delta Z$/$ \Delta R$\,=\,$-$0.034\,--\,0.043\,dex\,kpc$^{-1}$}.
    
    \item Spheroidal, disc and irregular galaxies at \zra{0.6{1.8}} have gas-phase metallicity gradients similar to each other. However we establish a strong correlation \changed{($r$\,=\,$-$1.0) at a 5\,--\,$\sigma$} level between the galaxies' stellar masses and metallicity gradients whereby more massive galaxies have steeper (more negative) abundance profiles 
    (Figure \ref{KGES_metal_fig:MHgrad_corr}). The rotation dominance (\vsig) is also identified to weakly correlate at a \changed{1.6\,--\,$\sigma$ level ($r$\,=\,$-$0.70)} with the metallicity gradients of our galaxies. More rotation-dominated galaxies have more negative metallicity gradients, as they are less efficient at redistributing metal-rich gas.
    
    \item Finally, we examine the metallicity gradients in the context of cosmic evolution (Figure \ref{KGES_metal_fig:MHgrad_z}). We demonstrate that our measurements of the metallicity gradients are consistent with flat gradients, that are reproduced in numerical simulations of high-redshift galaxies in which feedback plays a key role. We find no significant evolution between the two redshift samples with the \zra{0.6}{1.0} sample having a median metallicity gradient of \changed{\,$\Delta Z/\Delta R$\,=\,$-$0.004\,$\pm$\,0.005 dex kpc$^{-1}$ with a 16\,--\,84th percentile range of $-$0.03\,--\,0.05 dex kpc$^{-1}$}. At  \zra{1.2}{1.8} the median metallicity gradient is \changed{$ \Delta Z$/$ \Delta R$\,=\,$-$0.0002\,$\pm$\,0.00048 dex kpc$^{-1}$ with a 16\,--\,84th percentile range of $-$0.03\,--\,0.02 dex kpc$^{-1}$}.
\end{itemize}

Overall, we have shown that the gas-phase metallicity scaling relations of high-redshift star-forming galaxies are comparable to those in the local Universe, whilst being offset to lower metallicities. We found no difference between the metallicity gradients of irregular and disc galaxies. This may well be due to the limited spatial resolution of our observations, and the azimuthal smoothing that occurs when deriving a metallicity gradient. Future observations with VLT/ERIS and ELT/HARMONI will allow $\sim$kpc resolution observations of the interstellar medium of high-redshift galaxies, which will enable spatially-resolved scaling relations and metallicity gradients to be derived more accurately at high redshift.

\section*{Acknowledgements} We thank Richard Ellis and Alice Shapley for  useful discussions of this work and the referee for a constructive review. This work was supported by the Science and Technology Facilities Council (ST/L00075X/1). SG acknowledges the support of the Science and Technology Facilities Council through grant ST/N50404X/1 for support and the Cosmic Dawn Center of Excellence funded by the Danish National Research Foun-dation under then grant No. 140. ALT acknowledges support from a Forrest Research Foundation Fellowship, Science and Technology Facilities Council (STFC) grants ST/L00075X/1 and ST/P000541/1, and the ERC Advanced Grant DUSTYGAL (321334). MB acknowledges support from STFC rolling grants ‘Astrophysics at Oxford’ ST/H002456/1 and ST/K00106X/1. AJB acknowledges funding from the ``FirstGalaxies" Advanced Grant from the European Research Council (ERC) under the European Union's Horizon 2020 research and innovation programme (Grant agreement No. 789056). GEM acknowledges  the Villum Fonden research grant 13160 "Gas to stars, stars to dust: tracing star formation across cosmic time" and the Cosmic Dawn Center of Excellence funded by the Danish National Research Foun-dation under then grant No. 140. 

\section*{DATA AVAILABILITY}
The data underlying this article are available at  \url{http://astro.dur.ac.uk/KROSS/data.html} for the \zra{0.6}{1.0} sample. The \zra{1.2}{1.8} sample data will be made available at \url{http://astro.dur.ac.uk/kmos/kges/} upon the publication of Tiley et al. in prep.

%%%%%%%%%%%%%%%%%%%%%%%%%%%%%%%%%%%%%%%%%%%%%%%%%% 



\bibliographystyle{mnras}
\bibliography{master.bib} % if your bibtex file is called example.bib

%%%%%%%%%%%%%%%%%%%%%%%%%%%%%%%%%%%%%%%%%%%%%%%%%%

%%%%%%%%%%%%%%%%% APPENDICES %%%%%%%%%%%%%%%%%%%%%

\appendix
% \onecolumn
\FloatBarrier
\begin{figure*}
\section{Metallicity Profiles}\label{App:profile}
    \centering
    \includegraphics[scale=0.3]{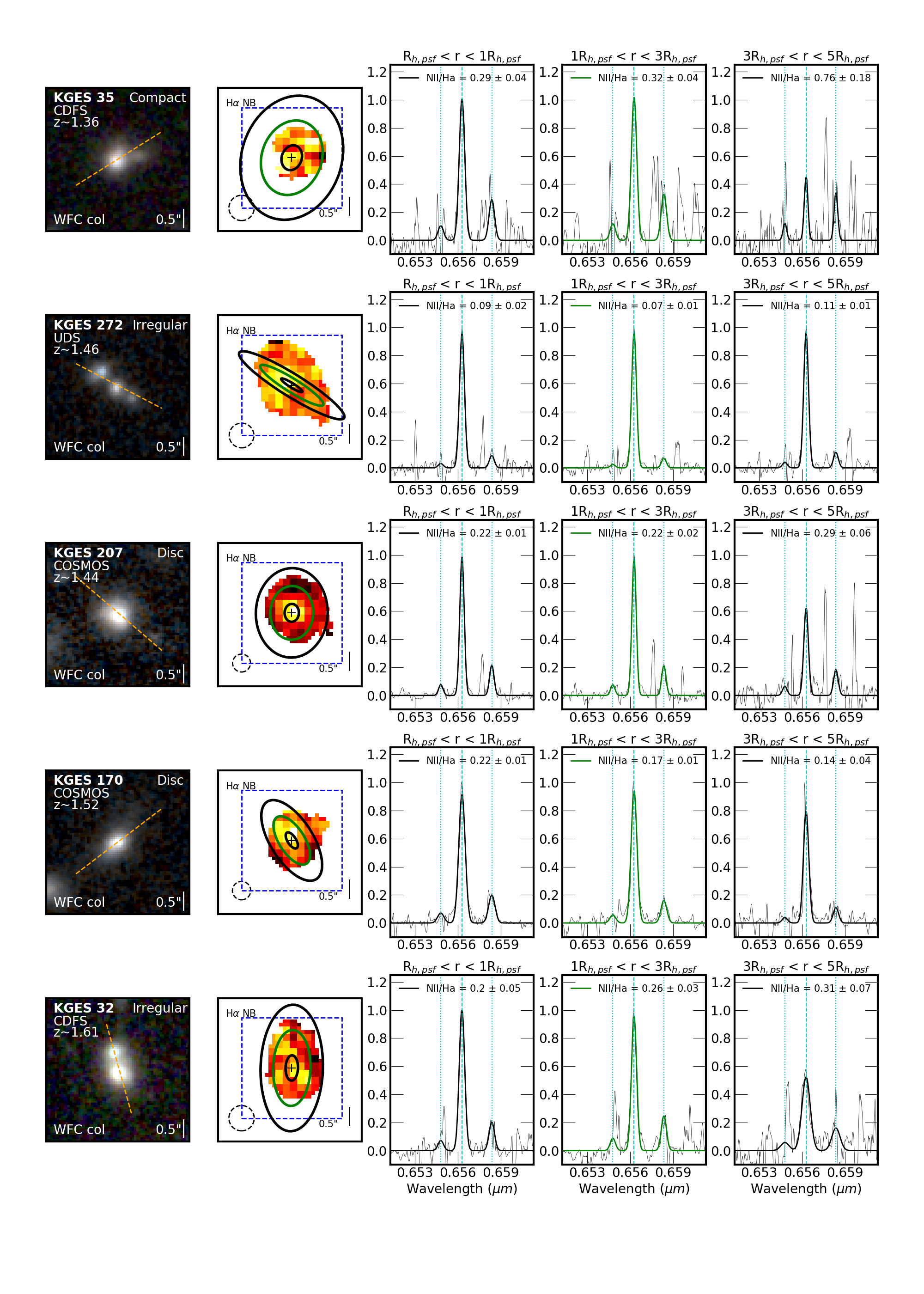}
    \caption[Examples of metallicity gradient derivation]{Examples of typical spatially-resolved galaxies, with from left to right, $HST$ colour image with the semi-major axis indicated (orange line). H$\alpha$ narrow band image from zero pointed observation, with KMOS field of view (blue dashed square) and annuli of multiples of $R_{\rm h,psf}$ (HWHM of a Gaussian PSF) with galaxy's axis ratio and position equal to the kinematic position angle of the galaxy. Spectrum extracted from each annulus is shown with the H$\alpha$ and  [{N\sc{ii}}] Gaussian model overlaid. A range of \NH{} profiles are shown with some galaxies displaying positive profiles increasing \NH{} line ratio with radius and others negative.}
     \label{KGES_metal_fig:NHratio_spec}
\end{figure*}

\onecolumn
\FloatBarrier
\section{Beam Smearing}\label{App:BS}
\begin{figure*}
    \centering
    \includegraphics[width=1\linewidth,trim={0cm 2cm 0cm 0cm}]{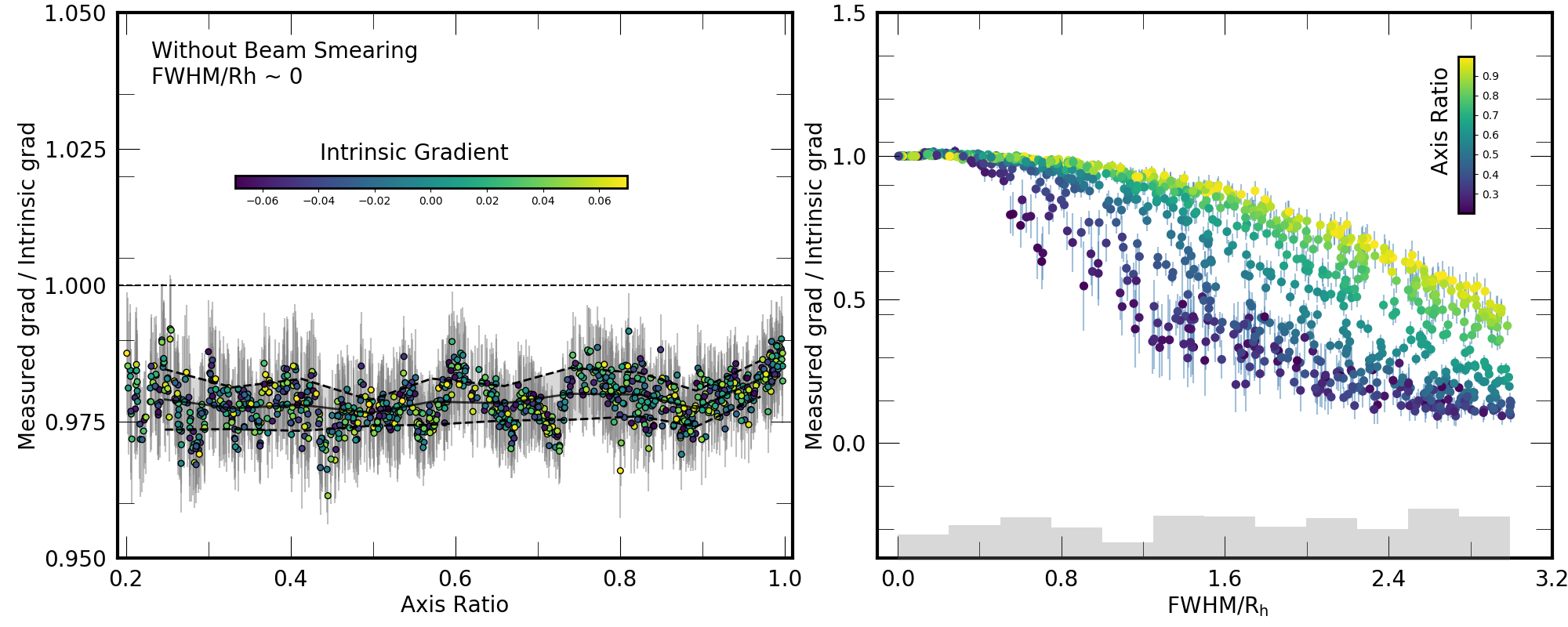}
    \caption[Beam smearing modelling at infinite S/N]{The ratio of measured metallicity gradient to intrinsic model gradient for 1000 mock galaxies of infinite signal to noise, as a function of axis ratio (left) and FWHM/\rh{} (right). We also show a histogram of FWHM/R$_{\rm h}$ values in the right hand panel. When the FWHM/\rh$\sim$\,0, we find no dependence on the axis ratio of the galaxy with a median of $(\Delta Z / \Delta R$ measured )/($ \Delta Z / \Delta R$ intrinsic )\,=\,0.98\,$\pm$\,0.01 with a scatter of 0.01. The offset from unity at all axis ratio is caused by the finite pixel sampling of the continuous distribution used to model the intrinsic gradient. When the FWHM/\rh$>$\,0, we identify a strong dependence on our ability to recover the intrinsic gradient as a function of FWHM/\rh{} as well as axis ratio. For a fixed FWHM/\rh{} ratio, the metallicity gradient in a more edge on galaxy less well recovered than a more face-on galaxy.}
    \label{KGES_metal_fig:model_grad_SNinf}
\end{figure*}
\begin{figure*}
    \centering
    \includegraphics[width=\linewidth,trim={0cm 4cm 0cm 0cm}]{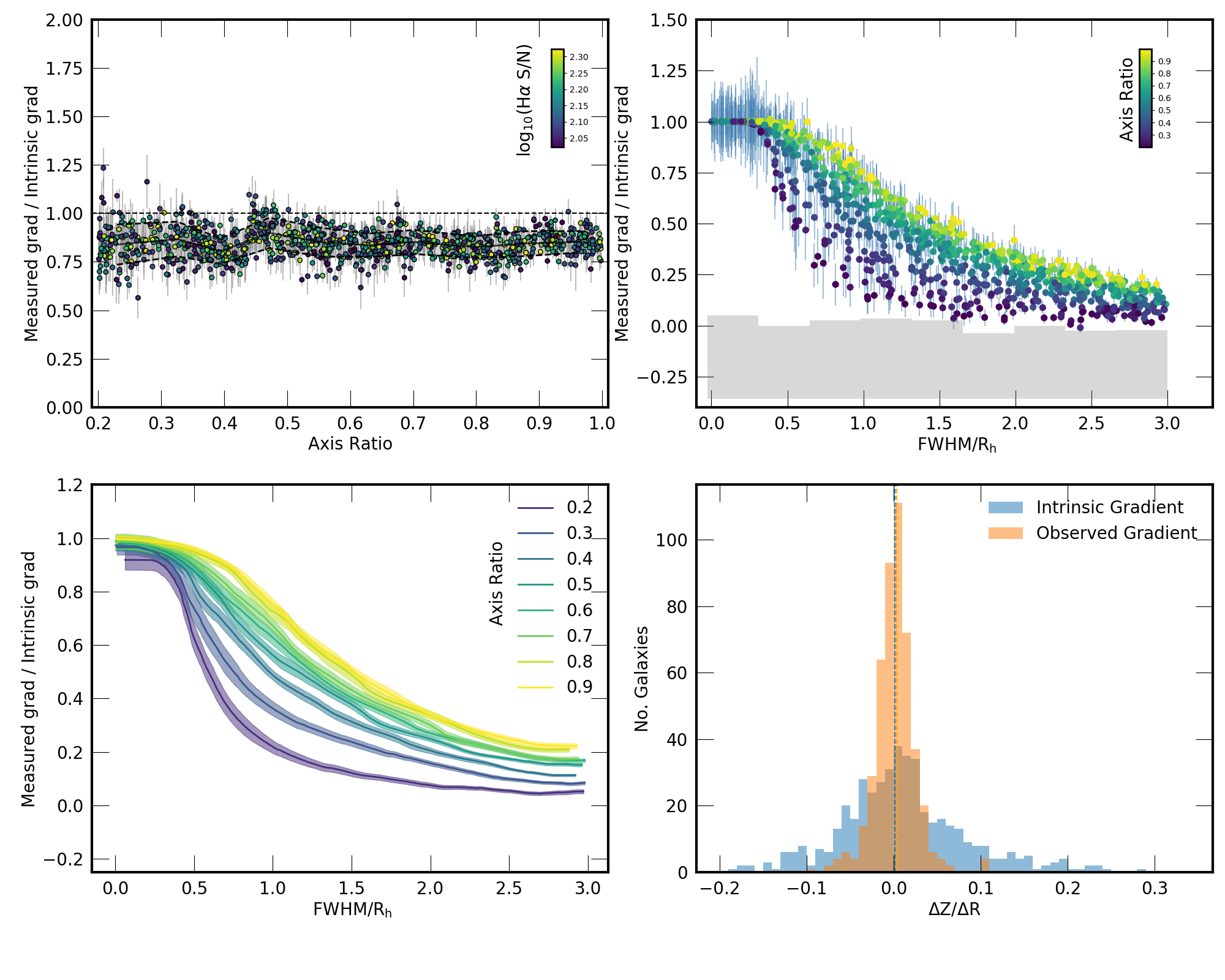}
    \caption[Beam smearing modelling at observable S/N]{The ratio of measured metallicity gradient to intrinsic model gradient for 1000 mock galaxies with signal to noise comparable to the KMOS observations, as a function of axis ratio (left) and FWHM/\rh{} (right). We also show a histogram of FWHM/\rh{} values in the right hand panel. When the FWHM/\rh{}$\sim$\,0, we find no dependence on the axis ratio of the galaxy with a median of  $(\Delta Z / \Delta R$ measured )/($\Delta Z / \Delta R$ intrinsic )\,=\,0.88\,$\pm$\,0.02 with a scatter of 0.2. The larger scatter is driven by the variation in signal to noise between the models. \changed{We apply this correction factor to the metallicity gradients before modelling the impact of beam smearing.} When the FWHM/\rh{}$>$\,0, we identify a strong dependence on our ability to recover the intrinsic gradient as a function of FWHM/\rh{} as well as axis ratio. For a fixed FWHM/R\rh{} ratio, the metallicity gradient in a more edge on galaxy is less well recovered than for a more face-on galaxy, however with more scatter than the infinite signal to noise version. The bottom left panel shows an interpolation of the ratio of the measured metallicity gradient to intrinsic model gradients as a function of FWHM/\rh{} at fixed axis ratio. \changed{The shaded regions show the 1\,--\,$\sigma$ uncertainty on each correction curve derived from bootstrapping the uncertainty on the derived metallicity gradient.} The bottom right panel shows a histogram of the observed metallicity gradient as well as the intrinsic, beam smearing corrected, metallicity gradient for the sample. The median observed metallicity gradient of the observational sample is \changed{$\Delta Z/ \Delta R$\,=\,0.0005\,$\pm$\,0.0013 dex kpc$^{-1}$ with a scatter of 0.037.} The median intrinsic metallicity gradient is \changed{$ \Delta Z/ \Delta R$\,=\,0.002\,$\pm$\,0.004 dex kpc$^{-1}$ with a scatter of 0.05.}}
    \label{KGES_metal_fig:model_grad_SN100}
\end{figure*}

\begin{figure}
    \centering
    \includegraphics[width=\linewidth]{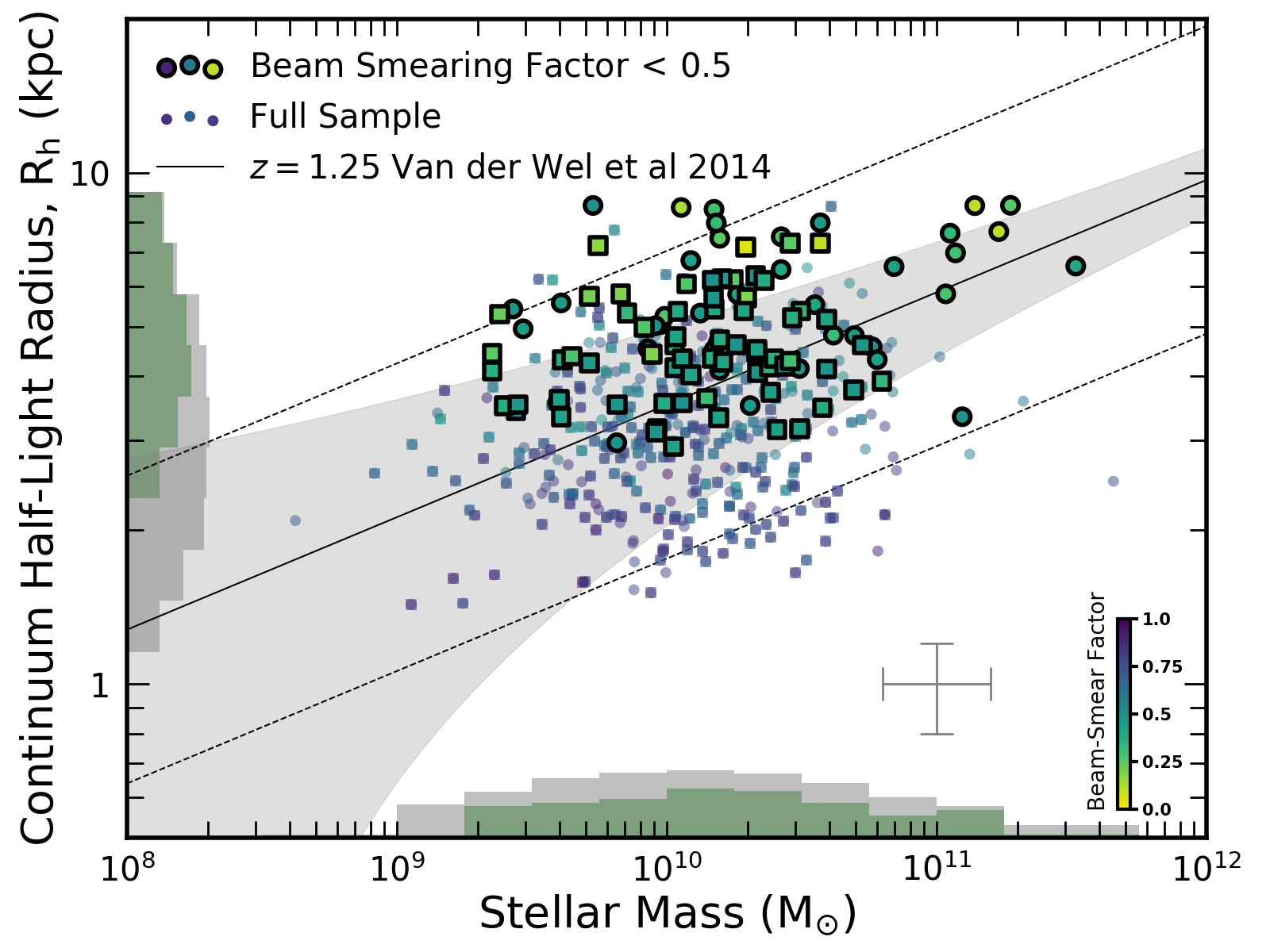}
    \caption{The stellar continuum half-light radius (\rh) as a function of stellar mass (M$_{*}$) for our sample at \zra{0.6}{1.0} ({\it{squares}})  and \zra{1.2}{1.8} ({\it{circles}}), coloured by the beam smearing correction factor derived for the galaxy. The median uncertainty on stellar mass and stellar continuum size are shown by the grey error bars in the lower right hand corner. We show histograms of both stellar mass and stellar continuum size on each axis. The solid line  (and shaded region) indicate the mass-size relation (and uncertainty) for star-forming galaxies at $z=1.25$ as derived by \citet{VanderWel2014}. The dashed lines are a factor of two above and below the relation. The \changed{98} galaxies with a beam-smearing correction factor to the metallicity gradient less than 50 per cent (black outlined points), have a similar stellar mass distribution to the overall sample whilst having larger continuum sizes, by definition.}
    \label{KMOS_metals:BS_sel}
\end{figure}

In this section we derive beam smearing corrections for the metallicity gradients of the galaxies.
As a first approach, to analyse the impact of axis ratio and beam smearing, we generate 1000 galaxies with infinite signal to noise. In Figure \ref{KGES_metal_fig:model_grad_SNinf} we show the ratio of the recovered metallicity gradient to the intrinsic gradient as a function of the galaxies axis ratio and the ratio of the half-light radius to the FWHM of the PSF. We identify no correlation with the axis ratio of the galaxy, in our ability to recover the intrinsic metallicity gradient of the galaxy. The offset from unity in Figure \ref{KGES_metal_fig:model_grad_SNinf}, is caused by the finite sampling on the pixel grid of the continuous S\'ersic distribution used to model the profile. The model galaxies have a median value of $(\Delta Z / \Delta R$ measured )/($\Delta Z / \Delta R$ intrinsic )\,=\,0.98\,$\pm$\,0.01. \changed{We correct for this finite sampling before modelling the impact of beam smearing on the metallicity gradients.}

In Figure \ref{KGES_metal_fig:model_grad_SNinf}  we also show the ratio of the measured gradient to intrinsic gradient as a function of FWHM/\rh{}. For galaxies with a larger FWHM/\rh{}, the accuracy of the metallicity gradient measurement reduces significantly. We also identify an axis ratio dependence, with more edge on galaxies at a fixed FWHM/\rh{} requiring a larger correction than face-on galaxies. A similar trend was found by \citet{Stott2014}. In more edge-on systems in the minor axis direction, the annuli are closer together, thus more affected by the spherical Gaussian PSF we use to model the turbulence in the atmosphere.

To understand the impact of signal to noise on our ability to measure the metallicity gradient, we generate a further 1000 galaxies with H$\alpha$ signal to noise comparable to that of the observations. In Figure \ref{KGES_metal_fig:model_grad_SN100} we show the relations between the ratio of measured and intrinsic gradient as a function of axis ratio and FWHM/\rh{}. We identify the same correlations with FWHM/\rh{} in the lower signal to noise models but with more scatter. The median beam smearing correction for galaxies with H$\alpha$ signal to noise comparable to the observations, across a range of FWHM/\rh{}, is $( \Delta Z/ \Delta R\, |_{\rm measured})$\,/\,$( \Delta Z/\Delta R\, |_{\rm intrinsic})$\,=\,0.45\,$\pm$\,0.02, i.e. the observed gradient is 45 per cent of the intrinsic gradient.

At fixed axis ratio, we interpolate the observed gradient to intrinsic gradient ratio as a function of FWHM/\rh{}. We use this correction curve \changed{(and uncertainty)} to derive beam smearing corrections for each galaxy in our sample. In Figure \ref{KGES_metal_fig:model_grad_SN100} we show a histogram of the observed and beam-smear correct intrinsic metallicity gradients of the sample. For a number of galaxies in our sample the beam smearing correction factor is larger than 50 per cent. A measurement of the intrinsic metallicity gradient of these galaxies is therefore very uncertain. We restrict our analysis of the metallicity gradients and their correlation with galaxy properties, to galaxies whose beam-smear correction factor is less than 50 per cent (\changed{98} galaxies). As shown in Figure \ref{KMOS_metals:BS_sel}, the sub-sample of \changed{98} galaxies have a similar distribution of stellar mass to the full observational sample whilst exhibiting larger stellar continuum sizes by this selection\hypertarget{lp}{.}

%%%%%%%%%%%%%%%%%%%%%%%%%%%%%%%%%%%%%%%%%%%%%%%%%%

% Don't change these lines
\bsp	% typesetting comment
\label{lastpage}
\end{document}